\documentclass[prd,nofootinbib,twocolumn]{revtex4}
\usepackage{graphicx, epsfig}
\usepackage{color}
\usepackage{mathrsfs}
\usepackage{bm}
\usepackage{amsmath,amssymb,yhmath}
\usepackage[caption=false]{subfig}
\usepackage{hyperref}
\usepackage[normalem]{ulem}
\usepackage{mathtools}
\usepackage{bigints}
\usepackage{float}
\definecolor{bittersweet}{rgb}{1.0, 0.44, 0.37}
\definecolor{coolblack}{rgb}{0.0, 0.18, 0.39}
\definecolor{britishracinggreen}{rgb}{0.0, 0.26, 0.15}
\definecolor{coolgrey}{rgb}{0.55, 0.57, 0.67}
\definecolor{darkgreen}{rgb}{0.0, 0.2, 0.13}
\definecolor{darkmagenta}{rgb}{0.55, 0.0, 0.55}
\definecolor{eggplant}{rgb}{0.38, 0.25, 0.32}
\definecolor{fashionfuchsia}{rgb}{0.96, 0.0, 0.63}

\newcommand{\be}{\begin{equation}}
\newcommand{\ee}{\end{equation}}
\newcommand{\ba}{\begin{eqnarray}}
\newcommand{\ea}{\end{eqnarray}}

\newcommand{\pa}{\partial}
\newcommand{\nn}{\nonumber}
\newcommand{\rmmax}{{\rm max}}
\newcommand{\la}{\langle}
\newcommand{\ra}{\rangle}

\newcommand{\half}{\frac{1}{2}}

\begin{document}
\title{Quantum Formation of Topological Defects}
\author{Mainak Mukhopadhyay$^{1}$}
\author{Tanmay Vachaspati$^{1}$}
\author{George Zahariade$^{1,2}$}
\affiliation{$^{1}$Physics Department, Arizona State University, Tempe, AZ 85287, USA. \\
}
\affiliation{$^{2}$Beyond: Center for Fundamental Concepts in Science, Arizona State University, Tempe, Arizona 85287, USA} 

\begin{abstract}
\noindent
We consider quantum phase transitions with global symmetry breakings that result in the formation of
topological defects. We evaluate the number densities of kinks, vortices, and monopoles that are produced
in $d=1,2,3$ spatial dimensions respectively and find that they scale as $t^{-d/2}$ and evolve towards attractor
solutions that are independent of the quench timescale. For $d=1$ our results apply in the region of 
parameters $\lambda \tau/m \ll 1$ where $\lambda$ is the quartic self-interaction of the order parameter, 
$\tau$ is the quench timescale, and $m$ the mass parameter.
\end{abstract}

\maketitle

\section{Introduction}
\label{introduction}

The formation of topological defects during a quantum phase transition is a novel 
process in which the quantum vacuum spontaneously breaks up into classical objects. 
In a thermal phase transition, the formation of defects is also a transition from a collection
of particles above the critical temperature to a collection of a complex of
particles (defects) and new excitations at low temperatures.
It is no surprise that there has been so much theoretical 
and experimental~\cite{Kibble:1976sj,Kibble:1980mv,Zurek:1985qw,Zurek:1993ek,Zurek:1996sj,
Zurek:2006dr,Vachaspati:2006zz,2007PhT....60i..47K,Chuang:1991zz,PhysRevLett.105.075701,
PhysRevLett.81.3703,Bowick:1992rz,Hendry1994,Ruutu1996,Bauerle1996,PhysRevLett.89.080603,
PhysRevLett.84.4966,PhysRevLett.91.197001,PhysRevLett.83.5210,ELTSOV20051,Beugnon2017,
PhysRevLett.96.180604,PhysRevB.74.144513} 
interest in understanding details of defect formation. 

The number density
of defects formed during a phase transition is sensitive to the rates at which external
parameters are changed to pass through the phase transition. The leading theoretical
framework for estimating the number density of defects is the ``Kibble-Zurek''
analysis~\cite{Kibble:1976sj,Kibble:1980mv,Zurek:1985qw,Zurek:1993ek,Zurek:1996sj,
Zurek:2006dr}. Numerical simulations have further strengthened the
model~\cite{Antunes:1998rz,Hindmarsh:2000kd,Yates:1998kx,Stephens:1998sm,Laguna:1996pv,
Donaire:2004gp,KOYAMA2006257}. However predictions of the Kibble-Zurek model have not 
yet gained universal confirmation, with most experiments in systems involving $^4$He, liquid 
crystals, superconductors, superfluids in
agreement~\cite{Chuang:1991zz,PhysRevLett.105.075701,PhysRevLett.81.3703,Bowick:1992rz,
Hendry1994,Ruutu1996,Bauerle1996,PhysRevLett.89.080603,PhysRevLett.84.4966,
PhysRevLett.91.197001,PhysRevLett.83.5210,ELTSOV20051,Beugnon2017,PhysRevLett.96.180604,
PhysRevB.74.144513} and others in disagreement~\cite{PhysRevLett.81.3703,
PhysRevB.60.7595,PhysRevLett.91.197001} with the predictions. In particular, the appearance 
of vortices in $^4$He was claimed in~\cite{Hendry1994} but was retracted 
in~\cite{PhysRevLett.81.3703} since it was found that the vortices in the former case were an 
externally induced artifact. Overall, the analysis of the phenomenon of defect formation in 
various systems is an ongoing field of research and has broad implications.

In the present work we follow our analysis of \cite{Mukhopadhyay:2020xmy} and solve for the number density of 
defects (kinks, vortices and monopoles)
formed during a quantum phase transition. The analysis is rigorous and without recourse to
approximation but the quantum field theory models we consider are ``free'', the only 
interaction being with external parameters that drive the phase transition. 
These models provide us with zeroth order solvable problems in different dimensions that 
we fully analyze.
Even with these minimal interactions, the analysis is highly non-trivial and in part has to be
done numerically. We discuss how other interactions may
be included in the analysis using perturbation theory and under what conditions we expect 
the zeroth order approximation to be accurate.

We are generally interested in Poincar\'e invariant field-theoretic models in 
$d+1$ spacetime dimensions, featuring an internal (global) $O(d)$ symmetry which is spontaneously 
broken during a quantum phase transition. In particular we will be considering $d$ real scalar fields 
$\Phi_{1},\dots,\Phi_{d}$ assembled in an $O(d)$-multiplet 
$\mathbf{\Phi}\equiv (\Phi_{1},\dots,\Phi_{d})^T$ whose dynamics are given by the Lagrangian 
density\footnote{In this paper we use a mostly plus signature for the Minkowski metric and 
natural units, $\hbar=c=1$.}
\be
{\cal L}^{(d)}=\frac{1}{2}\pa^\mu\mathbf{\Phi}^T\pa_\mu\mathbf{\Phi}
                                                 -V_{\beta}\left(\mathbf{\Phi}^T\mathbf{\Phi}\right)\,.
\label{Ld}
\ee
Here the potential $V_\beta$ is $O(d)$-invariant and depends on a (possibly time-dependent) 
external parameter $\beta$. We assume that $V_{\beta}$ is such that the vacuum manifold is 
$O(d)$-symmetric for $\beta<0$ and $O(d-1)$-symmetric for $\beta>0$. In other words, as the 
parameter $\beta$ increases from negative to positive values, the system transitions from a higher 
symmetry phase to a lower symmetry one, and the average vacuum field configuration starts 
exhibiting topological defects. These defects then annihilate with one 
another and eventually disappear. It is precisely this dynamics of formation and annihilation of 
topological defects that we are concerned with in this paper. In fact, our main purpose will be to 
determine the number density of topological defects as a function of time and its dependence
on the external parameter $\beta$,
using a combination of analytical and numerical methods.

\begin{figure}
      \includegraphics[width=0.45\textwidth,angle=0]{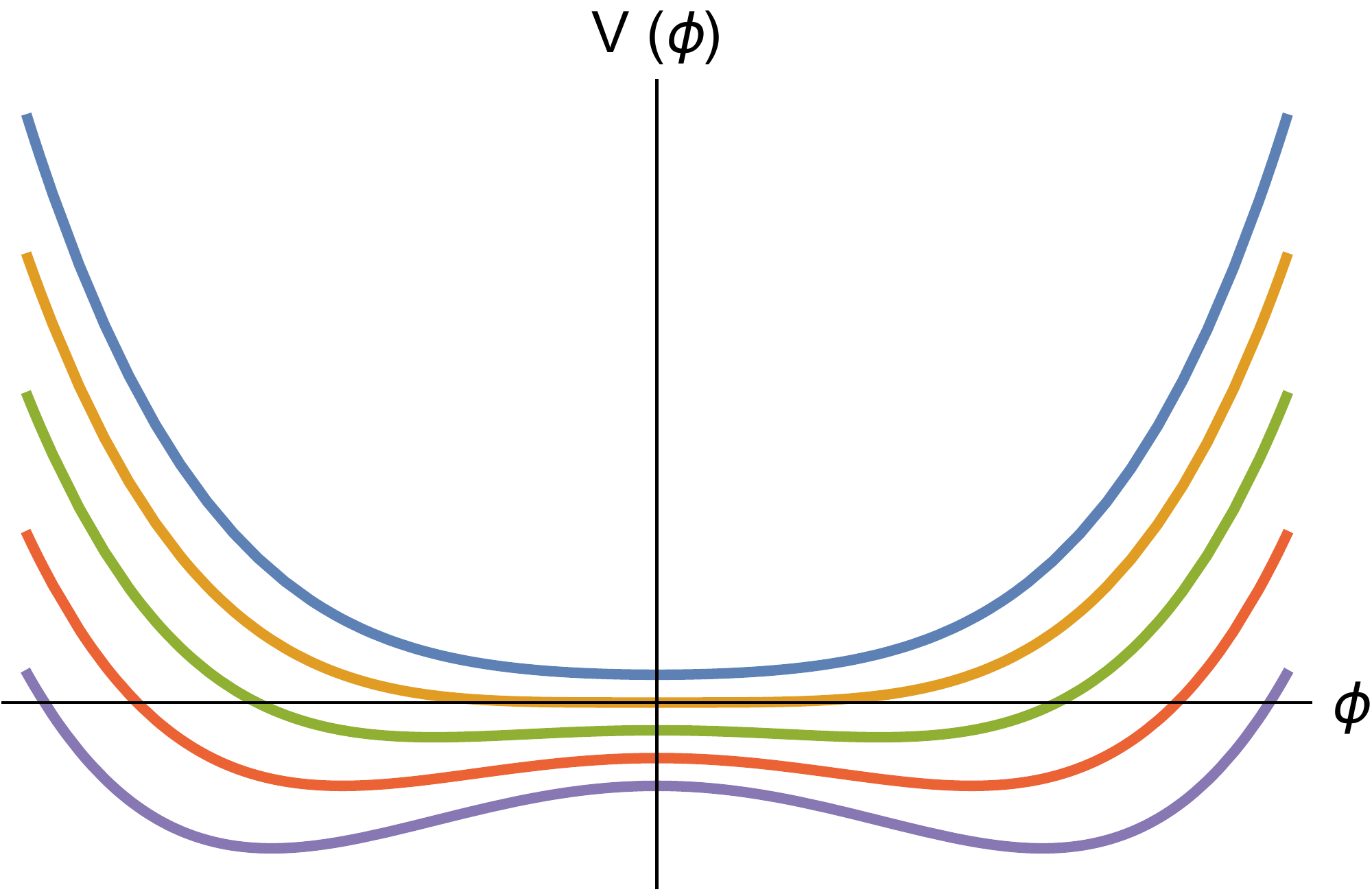}
 \caption{
Snapshots of the $d=1$ potential at a few different times. The plots have been
shifted vertically for clarity.
  }
\label{Vsnapshots}
\end{figure}

For concreteness we will take $\beta$ to be the (time-dependent) mass squared of the field, so that
\be
V_{\beta}\left(\mathbf{\Phi}^T\mathbf{\Phi}\right)
=\frac{1}{2}m_2(t)\mathbf{\Phi}^T\mathbf{\Phi}+\frac{\lambda}{4}\left(\mathbf{\Phi}^T\mathbf{\Phi}\right)^2\,,
\label{hypermexicanhat}
\ee
where 
\be
m_2(t)=-m^2\tanh\left(\frac{t}{\tau}\right)\,,
\label{m2t}
\ee
and $\lambda$, $m$, $\tau$ are positive parameters. In particular, the quench parameter $\tau$ is 
a time scale quantifying the rate of change of the potential during the phase transition. It is clear 
that for $t\ll -\tau$, the vacuum manifold reduces to the null field configuration $\Phi=0$
and is therefore $O(d)$-symmetric, while for 
$t\gg\tau$ it includes all field configurations on the $O(d-1)$-symmetric hypersphere given by
\be
\lambda\mathbf{\Phi}^T\mathbf{\Phi}=m^2\,.
\label{vac}
\ee
In Fig.~\ref{Vsnapshots} we sketch the potential~\eqref{hypermexicanhat} at a few different times.

It is well known that these models have topological defects -- kinks ($d=1$) in
one spatial dimension, vortices ($d=2$) in two spatial dimensions, and monopoles ($d=3$) in
three spatial dimensions~\cite{Vilenkin:2000jqa}. In each of these cases the vacuum manifold
described by \eqref{vac} has non-trivial topology: for $d=1$ it is 2 points, for $d=2$ it is
a circle, and for $d=3$ it is a two-sphere. The defect locations are described by
zeros of $\mathbf{\Phi}$ even in the symmetry broken phase. The zeros are trapped
due to the non-trivial topology of the vacuum manifold. We realize that the topology
persists even if we set $\lambda=0$ and the problem of defect formation simplifies.
Then the $\lambda=0$ problem can be thought of as the zeroth order problem.
We discuss the $\lambda \ne 0$ problem for $d=1$ in greater detail in Sec.~\ref{discussion}
where we find that $\lambda$ dependent corrections are small if 
$\lambda \tau /m \ll 1$.


The overall strategy will be to regulate both the IR and UV behaviors of the field theory by working in 
a finite box of size $L^d$ (with periodic boundary conditions) and discretize space on a $N^d$ point 
lattice (with lattice spacing $a=L/N$). Then we can determine an exact expression for the field probability 
density functional as a function of a finite number of quantities that can be computed numerically. We then 
find the average expectation value of a judiciously constructed quantum operator that counts the number 
density of zeros of the field multiplet $\mathbf{\Phi}$ in the limit of the finite resolution imposed by the lattice. 
We finally take both the continuum limit $N\rightarrow\infty$, $a\rightarrow 0$, and the infinite volume 
limit, $L\rightarrow\infty$ (in this exact order), to recover the full field theory result. Up to spurious zeros 
due to vacuum fluctuations that can consistently be discarded, this accurately gives the number density 
of topological defects. The case of a sudden phase transition ($\tau=0$) can be treated analytically
but the general case will be treated numerically. 

It should be mentioned that the so-called {\it spinodal decomposition} -- where one phase evolves into domains
of other phases in the absence of phase barriers -- during 
quantum phase transitions has been the subject of extensive work in the
 literature~\cite{calzetta1989spinodal,liu1991nonequilibrium,liu1992growth,liu1992defect,paufler,
PhysRevE.48.767,Boyanovsky:1999wd,Rivers:1995gk}. These studies were in the context of the 
Ginzburg-Landau model and in a more general field theoretic context but were limited to
instantaneous quenches~\cite{Ibaceta:1998yy,liu1991nonequilibrium,liu1992growth,liu1992defect,Rivers:1995gk}. Our purely quantum approach applies to non-instantaneous quenches 
and is readily generalizable to the case of $d$-dimensional 
global topological defects. The present work aims to describe it in an elementary and 
self-contained manner. We find, for different quench time-scales $\tau$, the behavior of the average defect
number density. We observe that defects start being produced immediately after the phase transition and their 
number density reaches a maximum within a short time, after which they start annihilating with each other and their 
number density goes down. The efficiency of topological defect production is found to depend on the details of the 
phase transition. Indeed the defect number density increases faster and to higher maximum values as $\tau$ 
decreases and the phase transition becomes more sudden. On the contrary, the late-time mutual annihilation of 
topological defects exhibits universal characteristics. After a transient regime, the number density of defects decays 
as a power law $t^{-d/2}$
with a coefficient that only depends on the spatial dimension $d$ and not on $\tau$.  Hence the $\tau=0$ 
result is an attractor for the dynamics of defect formation and subsequent decay for a large class of quantum phase 
transitions. Our comprehensive analysis thus provides a unifying picture of defect formation and decay during 
{\it non-instantaneous quenches} and fills a gap in the literature. We are however limited to the regime where the 
$\lambda=0$ approximation holds and we discuss this limitation in some detail in Sec.~\ref{discussion}.

The paper is structured as follows. In Sections~\ref{kinks} and~\ref{sec:vortices} we fully describe the 
average dynamics of kink ($d=1$) and vortex ($d=2$) condensation respectively. In Section~\ref{monopoles} 
we extend these results to 3 and higher dimensions. In Sec.~\ref{discussion} we discuss how the previous 
results constitute only the zeroth order approximation in a perturbative expansion in $\lambda$ and estimate 
the next-order corrections. We end in Section~\ref{conclusions} by emphasizing the importance of our results 
and contrasting them with previous work done on the subject.

\section{One Dimension: Kinks}
\label{kinks}

One of the challenges in finding the number density of kinks is to first define a kink in the
quantum field theory given by \eqref{Ld} with $d=1$, where we denote the single-component scalar field $\mathbf{\Phi}$ by $\phi$. This can be done using the
Mandelstam ``kink operator''~\cite{Mandelstam:1975hb}, which is a two-component 
{\it fermionic} operator, $\hat{\chi}$. A key property of $\hat{\chi}$ is that it satisfies the 
equal time commutation relations,
\be
\left[\hat{\phi}(t,y), \hat{\chi}(t,x)\right] = 
\begin{cases}
\eta \, \hat{\chi} (t,x), & y < x\,, \\
0, & y > x\,,
\end{cases}
\ee
where $\eta$ is a real number. If $|s\ra$ is an eigenstate of $\hat{\phi}(t,y)$ such that $\hat{\phi} |s\ra =0$ (for all $y$), then we find that the state
$|s'\ra \equiv \hat{\chi} (t,x) |s\ra$ satisfies
\be
\hat{\phi} (t,y) |s'\ra = 
\begin{cases}
\eta\, |s'\ra, & y < x \\
0, & y > x
\end{cases}
\ee
Hence the operator $\hat{\chi}$ has created a step in the value of $\phi$ at $x$ by an amount $\eta$.
If $\phi = 0$ and $\phi=\eta$ are two possible vacuum expectation values of $\phi$, $\hat{\chi}$ would 
have created a kink that interpolates between two vacua. The number density of $\chi$ quanta 
would then correspond to the number density of kinks.

Unfortunately the relation between $\chi$ and $\phi$ is quite complicated -- $\chi$ involves
exponentials of $\phi$ and ${\dot \phi}$ and other quantum field theory subtleties -- and we 
do not have a clear way to utilize the Mandelstam operator. Instead we find it useful to work 
entirely with the $\phi$ field, simply defining the kink to be a jump in the value of $\phi$ as further discussed in Sec.~\ref{subsec:kinks_avg}. Our definition of the kink operator
is also helpful in the case of vortices and monopoles for $d=2,3$ as these objects correspond to intersections of domain walls {\it i.e.} kinks extended to higher dimensions.

\subsection{Setup and quantization}
\label{subsec:setup}

We start by treating the $d=1$ case in detail. The relevant 
Lagrangian density for the real scalar field $\phi$ is thus
\be
\mathcal{L}^{(1)}= \half (\partial_\mu\phi)^2 - \half m_2(t) \phi^2 -\frac{\lambda}{4}\phi^4 \,.
\label{kinkqft}
\ee
Clearly, for $t<0$ the model has a unique vacuum $\phi=0$ while for $t>0$ it has two degenerate 
vacua at $\phi=\pm m/\sqrt{\lambda}$ corresponding to the two minima of the double-well potential. 
It is well-known that in the $t\gg\tau$ limit (where $m_2(t)\approx -m^2$), there exist static classical 
kink and anti-kink solutions given by
\be
\phi_{\pm} (x) = \pm  \frac{m}{\sqrt{\lambda}} \tanh \left ( \frac{m x}{\sqrt{2}}  \right )\,.
\ee
These solutions are non-perturbative and topologically non-trivial: they interpolate between the two 
vacua over a spatial scale $\sim 1/m$. Of course, Poincar\'e invariance allows the construction of 
displaced or even ``dynamical'' kinks from the above solutions but, whatever the frame, they will 
always be characterized by their topological charge
\be
q=\int_{-\infty}^{\infty} dx\, \pa_x \phi=\phi(\infty)-\phi(-\infty)\,.
\ee
In fact, a kink always has positive topological charge since the field undergoes a negative to positive 
sign change, while an anti-kink has the exact opposite property. 

Multi-kink and anti-kink solutions can be constructed as well, but these will not be static anymore 
since the kinks and anti-kinks will attract each other and they will eventually annihilate. If separations 
are large and the different kinks and anti-kinks are initially at rest, such configurations will however 
be approximately static. Even though the topological charge of such field configurations does not inform 
us about the number of kinks or anti-kinks involved (since the topological charge is a binary valued quantity), 
one can however in principle recognize the presence of individual kinks and anti-kinks in a given field 
configuration by focusing on the points where the field changes sign: a negative-to-positive sign change 
will be identified as a kink while a positive-to-negative one will be identified as an anti-kink. Of course this 
is only part of the picture because not every sign change should be counted as a kink or anti-kink especially 
if it occurs on time and distance scales shorter than the characteristic width of $1/m$. We will discuss this subtlety in Sec.~\ref{subsec:kinks_avg}.



We are interested in the production of kinks during a 
quantum phase transition and in particular in how their average number density scales with 
time. As we have discussed in Sec.~\ref{introduction}, we will first be analyzing the $\lambda=0$
case and the Lagrangian density we will work with will thus be
\be
\mathcal{L}^{(1)}= \half (\partial_\mu\phi)^2 - \half m_2(t) \phi^2\,.
\ee

We now need to quantize this model. We start by assuming that the volume (or length since $d=1$) of space is finite of size $L$ and that the field obeys periodic boundary conditions. (We can alternatively think of space as a circle of length $L$.) We then discretize space on a lattice consisting of $N$ points separated by a distance $a=L/N$. At each lattice point $x_j\equiv ja$, we define the discretized field $\phi_j\equiv \phi(x_j)$ and the full Lagrangian of the discretized theory reads
\be
\label{disc_lag}
L_{\rm disc.}^{(1)}=\frac{a}{2}\dot{\bm \phi}^T\dot{\bm \phi}- \frac{a}{2} {\bm \phi}^T\Omega_2(t){\bm \phi}\,,
\ee
where we have assembled the discretized fields in a column vector ${\bm \phi}\equiv({\phi}_1,\dots,{\phi}_N)^T$ and the matrix $\Omega_2$ is defined by
\be
[\Omega_2]_{jl} = 
\begin{cases}
+{2}/{a^2}+m_2(t)\,,& j=l\\
-{1}/{a^2}\,,& j=l\pm1\ (\text{mod}\ N)\\
0\,,&\text{otherwise}\,.
\end{cases}
\ee
Introducing the canonically conjugate momentum fields, $\pi_j\equiv a\dot{\phi}_j$, and assembling them 
in a column vector ${\bm \pi}\equiv({\pi}_1,\dots,{\pi}_N)^T$, we can promote both the $\phi_j$s and $\pi_j$s 
to operators satisfying canonical commutation relations $[\hat{\phi}_j,\hat{\pi}_l]=i\delta_{jl}$. The 
quantum Hamiltonian of the discretized theory~\cite{Vachaspati:2018hcu} then reads
\be
\hat{H}_{\rm disc.}^{(1)}=\frac{1}{2a}\hat{\bm \pi}^T\hat{\bm \pi}
+\frac{a}{2}\hat{\bm \phi}^T\Omega_2(t)\hat{\bm \phi}\,,
\label{hamdisc}
\ee
where hats denote operator valued quantities. It is apparent from~\eqref{hamdisc} that the discretized 
theory describes the quantum dynamics of a set of $N$ quadratically coupled, simple harmonic oscillators. 

We are interested in how the (unique) quantum vacuum before the phase transition (at a time $t_0\ll-\tau$ when the potential is upright and $m_2(t_0)\approx m^2$) is destabilized by the quench and evolves into a more complicated state featuring dynamical kinks and anti-kinks. To understand the dynamics of this process we need to solve the functional Schr\"odinger equation associated with~\eqref{hamdisc},
\be
\label{schrodinger}
i\frac{\pa \Psi}{\pa t} = -\frac{1}{2a} \Delta \Psi+\frac{a}{2}{\bm \phi}^T\Omega_2(t){\bm \phi}\ \Psi\,,
\ee
where the wave functional $\Psi[\phi_1,\dots,\phi_N;t]$ is such that $|\Psi|^2$ gives the probability density of a given field configuration at time $t$, and the Laplacian operator is defined by 
\be
\Delta\equiv \frac{\pa^2}{\pa\phi_1^2}+\dots+\frac{\pa^2}{\pa\phi_N^2}\,.
\ee
One can easily check that the wave functional for the vacuum state at $t=t_0$ is
\be
\Psi(t_0)=\mathcal{N}\exp\left[-\frac{a}{2}{\bm \phi}^T{\Omega_2(t_0)}^{1/2}\,{\bm \phi}\right],
\label{icwavefunc}
\ee
where 
\be
\mathcal{N}=\left(\frac{a}{\pi}\right)^{N/4}{\rm det}\left(\Omega_2(t_0)\right)^{1/8}\,,
\ee
and fractional powers of the positive-definite matrix $\Omega_2(t_0)$ 
are unambiguously defined in the standard way. For instance $\Omega_2^{1/2}=O\,\text{Diag}\left(\lambda_1^{1/2},\dots,\lambda_N^{1/2}\right)O^T$, where $O$ is the orthogonal matrix diagonalizing $\Omega_2$, and $\lambda_j$ are the (positive) eigenvalues of $\Omega_2$.
Given this initial condition, the solution for the wave functional at time $t$ will be given by
\be
\label{wavefunctional}
\Psi (t) = 
\mathcal{N}
\exp \left [ -\half \int_{t_0}^t dt' {\rm Tr}\, M(t')+ \frac{ia}{2} {\bm \phi}^T M(t)\, {\bm \phi} \right ],
\
\ee
where the $N\times N$ complex symmetric matrix $M(t)$ verifies 
\be
\label{Meqn}
\dot{M}+M^2+\Omega_2(t)=0\,,
\ee
and $M(t_0)=i\Omega_2(t_0)^{1/2}$.
Introducing the complex $N\times N$ matrix $Z(t)$ defined by
\be
\ddot{Z}+\Omega_2(t)Z=0\,,
\label{Zeq}
\ee
and
\ba
Z(t_0)&=&-\frac{i}{\sqrt{2a}}\Omega_2(t_0)^{-1/4}\,,
\label{Zic}\\
\dot{Z}(t_0)&=&\frac{1}{\sqrt{2a}}\Omega_2(t_0)^{1/4}\,,
\label{Zdotic}
\ea
we can write 
\be
M=\dot{Z}Z^{-1}\,.
\ee
Indeed, using~\eqref{Zeq}, \eqref{Zic} and~\eqref{Zdotic}, it is easy to check that this expression 
yields a symmetric matrix since $\dot{Z}Z^{-1}-(\dot{Z}Z^{-1})^T$ is a conserved quantity which 
vanishes at time $t_0$. We can 
now write the probability density functional as
\ba
|\Psi(t)|^2=|\mathcal{N}|^2\exp &\Biggr[& -\frac{1}{2}\int_{t_0}^tdt'{\rm Tr}\left(M(t')+M(t')^\dag\right)\nn\\
&&+\frac{ia}{2} {\bm \phi}^T \left(M(t)-M(t)^\dag\right)\, {\bm \phi} \Biggr]\,.
\label{probdens}
\ea
To simplify this expression we first use the fact that
\be
\int_{t_0}^tdt'{\rm Tr}\left(M(t')+M(t')^\dag\right)=\left.{\rm Tr}\left(\log K\right)\right|^t_{t_0}\,,
\label{dettrlog}
\ee
where $K\equiv ZZ^\dag$ is a real positive definite symmetric matrix; indeed, using~\eqref{Zeq}, 
\eqref{Zic} and~\eqref{Zdotic}, it is easy to check that $ZZ^\dag-Z^*Z^T$ is a conserved quantity 
which vanishes at time $t_0$.
Next, we make use of another conserved quantity 
\be
Z^\dag\dot{Z}-\dot{Z}^\dag Z=i/a\,,
\label{angular}
\ee 
which can also be verified via~\eqref{Zeq}, \eqref{Zic} and~\eqref{Zdotic}, to show that
\be
M(t)-M(t)^\dag=iK^{-1}/a\,.
\label{covariance}
\ee
Finally, plugging~\eqref{dettrlog} and~\eqref{covariance} into~\eqref{probdens} yields a simplified 
(and manifestly normalized) expression for the probability density functional
\be
|\Psi(t)|^2= \frac{1}{\sqrt{{\rm det}(2\pi K})} e^{- {\bm \phi}^T K^{-1} {\bm \phi} / 2}\,.
\label{probdensfin}
\ee
This expression (along with~\eqref{Zeq}, \eqref{Zic} and~\eqref{Zdotic}) contains all the information that 
we will need in order to determine the average number density of kinks in the lattice. Note that $K$ is 
a time-dependent matrix, whose time-dependence is given by that of the matrix $Z$. 

Before going any further, we mention a separate interpretation of the matrix $Z$. Working in Heisenberg 
picture with respect to time $t_0$, we can define creation and annihilation operators at time $t_0$ by
\ba
\hat{\bm a}(t_0)&\equiv&\frac{1}{\sqrt{2a}}\left(\Omega_2^{-1/4}\hat{\bm \pi}(t_0)
                                         -ia\Omega_2^{1/4}\hat{\bm \phi}(t_0)\right)\,, \\
\hat{\bm a}^\dag(t_0)&\equiv&\frac{1}{\sqrt{2a}}\left(\Omega_2^{-1/4}\hat{\bm \pi}(t_0)
                                              +ia\Omega_2^{1/4}\hat{\bm \phi}(t_0)\right) \,.
\label{create_annhil}
\ea
Notice that we have used column vector notation here but that the dagger refers to the adjoint operation 
on the Hilbert space only: it does not turn column vectors into row vectors. Then we can expand the 
Heisenberg picture discretized field operators at time $t$ as follows
\be
\hat{\bm \phi}(t)=Z(t)^*\hat{\bm a}(t_0)+Z(t)\hat{\bm a}^\dag(t_0)\,.
\label{heisenbergexp}
\ee
Eqs.~\eqref{Zeq}, \eqref{Zic} and~\eqref{Zdotic} ensure that the Heisenberg equations as well as the 
proper initial conditions at $t_0$ are verified. Now it is easy to see that the matrix $K$ is simply the 
covariance matrix of the discretized field since, using~\eqref{heisenbergexp},
\be
\langle 0|\hat{\phi}_j\hat{\phi}_l|0\rangle = \sum_{k=1}^NZ^*_{jk}Z_{lk}=K_{lj}\,.
\label{phit}
\ee
Here the Heisenberg picture vacuum $|0\rangle$ is time-independent and defined by the 
wave functional~\eqref{icwavefunc}.

In principle we now have all the ingredients needed to discuss the quantum production of kinks 
during the phase transition. Indeed equation~\eqref{probdensfin} along with the $N^2$ complex 
linear ordinary differential equations~\eqref{Zeq} fully determine the quantum dynamics of the field 
configuration. However, it turns out that not all components of the matrix $Z$ are relevant and we 
can reduce the number of differential equations that need to be solved. It can be 
shown that the matrix $Z$ is {\it circulant}~\cite{Mukhopadhyay:2019hnb} i.e., its matrix elements $Z_{jl}$ only 
depend on $j-l\ (\text{mod}\ N)$. We can therefore diagonalize it via the discrete Fourier transform:
\be
Z_{jl}=  \frac{1}{\sqrt{N}} \sum_{n=1}^N c_n(t) e^{-i(j-l)2\pi n /N}\,.
\label{Zceq}
\ee
This allows us to recast~\eqref{Zeq}, \eqref{Zic} and~\eqref{Zdotic} in terms of the complex mode 
functions $c_n(t)$ thus obtaining
\be
{\ddot c}_n + \left [ \frac{4}{a^2}\sin^2 \left (\frac{\pi n}{N} \right ) + m_2(t) \right ] c_n = 0\,,
\label{ckeq}
\ee
and 
\be
c_n (t_0) = 
\frac{-i}{\sqrt{2L}} 
\left [ \frac{4}{a^2}\sin^2 \left ( \frac{\pi n}{N}\right ) + m_2(t_0) \right ]^{-1/4}\,,
\label{ckt0}
\ee
\be
{\dot c}_n (t_0) = \frac{1}{\sqrt{2L}} 
\left [ \frac{4}{a^2}\sin^2 \left (\frac{\pi n}{N} \right ) + m_2(t_0) \right ]^{1/4}\,.
\label{dotckt0}
\ee
Rewriting the dynamical equations in terms of mode coefficients provides an enormous computational 
gain: we now only have to solve $N$ equations instead of $N^2$. Additionally, as we will shortly see, 
mode coefficients are particularly well suited to discussing problems related to the $N\rightarrow\infty$ 
limit and divergences related to vacuum fluctuations of the quantum field. We can achieve further 
simplification by writing the mode functions in trigonometric form 
\be
c_n\equiv\rho_n e^{i\theta_n}\,,
\ee
where $\rho_n$ and $\theta_n$ are respectively the modulus and argument of the complex number 
$c_n$, and making use of the conserved quantity~\eqref{angular}, which in this representation takes 
the form of a conserved angular momentum,
\be
\rho_n^2\dot{\theta}_n=1/2L\,.
\ee
Then~\eqref{ckeq} reduces to a set of $N$ real (but non-linear) ordinary differential equations,
\be
\label{rhok}
\ddot{\rho}_n +\left [ \frac{4}{a^2}\sin^2 \left (\frac{\pi n}{N} \right ) + m_2(t) \right ] \rho_n 
  = \frac{1}{4L^2\rho_n^3}\,,
\ee
with initial conditions
\ba
\label{rhokic}
\rho_n(t_0)&=&\frac{1}{\sqrt{2L}} 
\left [ \frac{4}{a^2}\sin^2 \left ( \frac{\pi n}{N}\right ) + m_2(t_0) \right ]^{-1/4}\,,\\
\dot{\rho}_n(t_0)&=&0\,.
\label{rhokdotic}
\ea

Even though working in terms of modes is computationally advantageous, kinks are
configurations (field zeros) in physical space. Thus we have to straddle the two descriptions
as in the following Sec.~\ref{subsec:kinks_avg}.

\subsection{Average kink number density}
\label{subsec:kinks_avg}

We are now in a position to tackle the problem of kink production during the phase transition. 
As mentioned earlier, since kinks and anti-kinks occur at zeros of the field configuration we first 
introduce a quantum operator $\hat{n}_Z$ that gives the number density of zeros in a given field 
configuration:
\ba
\hat{n}_Z = n_Z\left(\hat{\bm\phi}\right) &\equiv& \frac{1}{L} \sum_{j=1}^N \frac{1}{4} \left [
\text{sgn}\left(\hat{\phi}_j\right) - \text{sgn}\left(\hat{\phi}_{j+1}\right) \right ]^2\nn\\
&=&\frac{N}{2L}-\frac{1}{2L}\sum_{j=1}^N\text{sgn}\left(\hat{\phi}_j\hat{\phi}_{j+1}\right)\,.
\ea
More precisely, such an operator is sensitive to the number of sign changes that occur between 
adjacent points on the lattice. We should stress that this is only accurate up to the finite resolution 
given by the lattice spacing $a$. It may in fact undercount the number of zeros of the actual continuous 
field configuration (if there are multiple sign changes within a lattice spacing). We expect however that, 
as $N$ becomes large enough, this operator will become more and more accurate. This assumption 
is reasonable as long as we can find a way to disregard high frequency noise-like fluctuations due 
to the quantum vacuum thus only counting ``true'' kinks and anti-kinks.

 We now calculate the vacuum expectation value of this operator or, in Heisenberg picture
\be
\langle \hat{n}_Z\rangle \equiv \langle 0|\hat{n}_Z(t)|0\rangle\,. 
\ee
Given that we know the probability density functional explicitly for the Schr\"odinger picture 
time-dependent state we can write
\ba
\langle \hat{n}_Z\rangle &=& \frac{1}{\sqrt{{\rm det}(2\pi K})} 
\int d^N{\bm \phi}\ n_Z(\bm \phi) e^{- {\bm \phi}^T K^{-1} {\bm \phi} / 2}\nn\\
&=& \frac{N}{2L}-\frac{1}{2L}\sum_{j=1}^N\left\langle{\rm sgn}\left(\hat{\phi}_j\hat{\phi}_{j+1}\right)\right\rangle \,,
\ea
where
\ba
\left\langle{\rm sgn}\left(\hat{\phi}_j\hat{\phi}_{j+1}\right)\right\rangle 
                   &\equiv& \frac{1}{\sqrt{{\rm det}(2\pi K)}}\times\nn\\
&&\hspace{-1cm}\int d^N{\bm \phi}\ \text{sgn}\left(\phi_j\phi_{j+1}\right) e^{- {\bm \phi}^T K^{-1} {\bm \phi} / 2}\,.
\label{inteq}
\ea
Introducing the permutation (shift) matrix 
\be
P_{ij}=
\begin{cases}
1\,,&j=i+1\, (\text{mod}\ N)\\
0\,,&\text{otherwise}\,,
\end{cases}
\ee
and performing the change of variables ${\bm \phi}\rightarrow P^{1-j}{\bm \phi}$, we can rewrite~\eqref{inteq} as
\ba
\left\langle{\rm sgn}\left(\hat{\phi}_j\hat{\phi}_{j+1}\right)\right\rangle &=& \frac{1}{\sqrt{{\rm det}(2\pi K)}}\times\nn\\
&&\hspace{-1.7cm}\int d^N{\bm \phi}\ \text{sgn}\left(\phi_1\phi_{2}\right) e^{- {\bm \phi}^TP^{j-1} K^{-1} P^{1-j}{\bm \phi} / 2}\,.
\ea
As mentioned earlier, $Z$ is a circulant matrix and, consequently, it has to be 
polynomial in $P$. Therefore, the matrix $K=ZZ^\dag$ is also circulant and $K^{-1}$ is seen 
to commute with $P$. This implies that,
\ba
\left\langle{\rm sgn}\left(\hat{\phi}_j\hat{\phi}_{j+1}\right)\right\rangle 
&=&\left\langle{\rm sgn}\left(\hat{\phi}_1\hat{\phi}_{2}\right)\right\rangle=\nn\\
&&\hspace{-3cm} \frac{1}{\sqrt{{\rm det}(2\pi K)}}\int d^N{\bm \phi}\ \text{sgn}\left(\phi_1\phi_{2}\right) 
e^{- {\bm \phi}^T K^{-1} {\bm \phi} / 2} \,,
\label{integral}
\ea
and the average number density of zeros simply reduces to
\be
\langle \hat{n}_Z\rangle 
= \frac{N}{2L}\left[1-\left\langle{\rm sgn}\left(\hat{\phi}_1\hat{\phi}_{2}\right)\right\rangle \right]\,.
\label{nZsgn}
\ee
Let it be mentioned here that the circulant property of the covariance matrix $K$ is the mathematical 
counterpart of the fact that the system has translational invariance (which is maintained at a discretized 
level by our choice of periodic boundary conditions). In other words, it is a consequence of the fact that 
two-point correlation functions $\langle\phi(x)\phi(y)\rangle$ only depend on the relative position $|x-y|$. 

We now need to evaluate~\eqref{integral} more explicitly. We start by writing
\ba
\label{intquad}
\left\langle{\rm sgn}\left(\hat{\phi}_1\hat{\phi}_{2}\right)\right\rangle
&=&\frac{1}{\sqrt{{\rm det}(2\pi K)}}\times\\
&&\hspace{-2.5cm}\sum_{Q=I}^{IV}
\int_Q d\phi_1d\phi_2\ {\rm sgn}(\phi_1\phi_2)\int d\phi_3\dots d\phi_N\, 
e^{- {\bm \phi}^T K^{-1} {\bm \phi}/2}\nn\,,
\ea
where the sum runs over the four quadrants in the $(\phi_1,\phi_2)$ plane (denoted by Roman numerals). 
We then decompose $K^{-1}$ into suitably sized blocks,
\be
K^{-1}  = (ZZ^\dag )^{-1} =\begin{pmatrix} A & B \\ B^T & C \end{pmatrix}\,,
\ee
where $A$ and $C$ are real
symmetric matrices of respective sizes $2\times 2$ and $(N-2)\times(N-2)$, while $B$ is a 
$2\times(N-2)$ real matrix, and introduce the notations ${\bm \chi} = (\phi_1,\phi_2)^T$, 
${\bm \xi}=(\phi_3,\dots,\phi_N)^T$. We also assume that $C$ is invertible, which will be true generically.
This allows us to rewrite the bilinear in the exponent in~\eqref{intquad} as
\ba
{\bm \phi}^T K^{-1} {\bm \phi}&=& ({\bm\xi} + C^{-1} B^T {\bm \chi}  )^T C ({\bm \xi} + C^{-1} B^T {\bm\chi} )  \nn\\
&& \hspace{1cm} +{\bm \chi}^T (A-BC^{-1}B^T) {\bm\chi} \,.
\ea
Using
\be
\int d^{N-2}{\bm \xi}\ e^{- ({\bm \xi} + C^{-1} B^T {\bm \chi}  )^T C ({\bm \xi} + C^{-1} B^T {\bm \chi} ) /2} 
= 
\frac{(2\pi)^{(N-2)/2}}{ \sqrt{{\rm det}(C)}}\,,
\ee
we can perform the Gaussian integral over $\phi_3, \dots,\phi_N$ and obtain
\ba
\left\langle{\rm sgn}\left(\hat{\phi}_1\hat{\phi}_{2}\right)\right\rangle&=&
\frac{1}{2\pi\sqrt{{\rm det}(K){\rm det}(C)}}\times\label{sgneq}\\
&&\hspace{-2cm}\sum_{Q=I}^{IV}\int_Q d\phi_1d\phi_2\ {\rm sgn}(\phi_1\phi_2)\,\exp\left[-\frac{1}{2} 
( \phi_1, \phi_2) A' \begin{pmatrix} \phi_1 \\ \phi_2 \end{pmatrix}\right],\nn
\ea
where 
\be
A'\equiv A-BC^{-1}B^T
\ee 
is the so-called {\it Schur complement} of $C$. The left-over two-dimensional quadrant integrals 
\be
I_Q\equiv \int_Q d\phi_1d\phi_2\ {\rm sgn}(\phi_1\phi_2)\,
\exp\left[ -\frac{1}{2} 
( \phi_1, \phi_2)  A'  \begin{pmatrix} \phi_1 \\ \phi_2 \end{pmatrix}\right],
\ee
can also be carried out. For the first quadrant, for example, ${\rm sgn}(\phi_1\phi_2)=+1$ and we can write
\ba
I_{I}&=&\int_0^{\infty}\int_0^\infty d\phi_1d\phi_2\times \nn\\
&&\quad\exp\left[ -\frac{1}{2}\left(A'_{11}\phi_1^2 +2A'_{12}\phi_1\phi_2+A'_{22}\phi^2_2\right)\right]\nn\\
&=&\int_0^{\infty}ds \int_0^\infty \phi_2\, d\phi_2 \times \nn\\
&&\quad\exp\left[ -\frac{1}{2}\left(A'_{11}s^2 +2A'_{12}s+A'_{22}\right)\phi_2^2\right]\nn\\
&=&\int_0^\infty ds\ \frac{1}{A'_{11}s^2 +2A'_{12}s+A'_{22}}\nn\\
&=&\frac{1}{\sqrt{A'_{11}A'_{22}-A'_{12}}}\left[\frac{\pi}{2}-\tan^{-1}\left(\frac{A'_{12}}{\sqrt{A'_{11}A'_{22}-A'_{12}}}\right)\right]\nn\\
&=&\frac{1}{\sqrt{{\rm det}(A')}}\left[\frac{\pi}{2}-\tan^{-1}\left(\frac{A'_{12}}{\sqrt{{\rm det}(A')}}\right)\right],
\label{bigint}
\ea
where in going from the first to the second line we used the change of variables $\phi_1\rightarrow s\phi_2$. 

The integrals over the remaining three quadrants
are readily obtained from $I_I$ as follows. To begin with, the change of variables $\phi_1\rightarrow-\phi_1$ and $\phi_2\rightarrow-\phi_2$ makes it clear that $I_{III}=I_{I}$, and $I_{II}=I_{IV}$. Furthermore, notice that the change of variables $\phi_1\rightarrow -\phi_1$ (leaving $\phi_2$ unchanged) on $I_{II}$ has the same effect (up to an overall sign) as changing $A'_{12}$ into $-A'_{12}$ in~\eqref{bigint}. We thus obtain
\be
I_{II}=-\frac{1}{\sqrt{{\rm det}(A')}}\left[\frac{\pi}{2}+\tan^{-1}\left(\frac{A'_{12}}{\sqrt{{\rm det}(A')}}\right)\right],
\ee
and all the four integrals $I_Q$ appearing in~\eqref{sgneq} are accounted for. We can achieve further simplification by taking advantage of the properties of the matrix $A'$. In particular since
\be
K^{-1}=
\begin{pmatrix}
I & BC^{-1} \\
0 & I
\end{pmatrix}
\begin{pmatrix}
A' & 0 \\
0 & C
\end{pmatrix}
\begin{pmatrix}
I & 0 \\
C^{-1}B^T & I
\end{pmatrix},
\label{mateq}
\ee
we have 
\be
\label{detk_inv}
{\rm det}\left(K^{-1}\right)=\frac{1}{{\rm det}\left(K\right)}={\rm det}\left(A'\right){\rm det}\left(C\right)
\ee
and~\eqref{sgneq} collapses to
\be
\left\langle{\rm sgn}\left(\hat{\phi}_1\hat{\phi}_{2}\right)\right\rangle
=-\frac{2}{\pi}\tan^{-1}\left(\frac{A'_{12}}{\sqrt{{\rm det}(A')}}\right).
\label{sgneq2}
\ee
But we can go even further. Indeed inverting~\eqref{mateq},
\be
\label{kmatrix}
K=
\begin{pmatrix}
I & 0 \\
-C^{-1}B^T & I
\end{pmatrix}
\begin{pmatrix}
A'^{-1} & 0 \\
0 & C^{-1}
\end{pmatrix}
\begin{pmatrix}
I & -BC^{-1} \\
0 & I
\end{pmatrix},
\ee
shows that $A'^{-1}$ coincides with the upper-left $2\times 2$ block of the matrix $K$. 
More explicitly we can write,
\be
A'^{-1}=\begin{pmatrix}
\alpha & \beta \\
 \beta & \alpha
\end{pmatrix}
\ee
where, using~\eqref{Zceq} and the reality of $K$,
\ba
\alpha &\equiv& K_{11} = \sum_{n=1}^N |c_n|^2 ,\label{alpha} \\
\beta &\equiv& K_{12} = \sum_{n=1}^N |c_n|^2 \cos (2\pi n/N)\,.\label{beta}
\ea
Thus
\be
A'=\frac{1}{\alpha^2-\beta^2}\begin{pmatrix}
\alpha & -\beta \\
 -\beta & \alpha
\end{pmatrix}
\ee
and~\eqref{sgneq2} becomes
\ba
\left\langle{\rm sgn}\left(\hat{\phi}_1\hat{\phi}_{2}\right)\right\rangle
&=&\frac{2}{\pi}\tan^{-1}\left(\frac{\beta}{\sqrt{\alpha^2-\beta^2}}\right)\nn\\
&=&\frac{2}{\pi}\sin^{-1}\left(\frac{\beta}{\alpha}\right)\,.
\ea
Finally we obtain the average number density of zeros
\be
\langle \hat{n}_Z\rangle = \frac{N}{2L}\left[1-\frac{2}{\pi} \sin^{-1}\left(\frac{\beta}{\alpha}\right)\right]\,.
\label{finalxnz}
\ee

Recall however that we are interested in the average number density of kinks which may
differ from the number density of zeros as given in \eqref{finalxnz} because the latter includes
zeros due to vacuum fluctuations of the quantum field. 
The difference between the two quantities is most clear long before 
the phase transition, where the field is in its unique vacuum and its expectation value vanishes 
everywhere on the lattice. However the field fluctuates about zero and there is
a non-zero average number density of zeros. This is to be contrasted with 
the average number density of kinks which should be exactly zero before the phase transition. 
Moreover the average number density of zeros is expected to be highly sensitive to the number of 
lattice points $N$ since the finer the resolution, the more zeros can be identified. This is again 
different for the average number density of kinks which are supposed to be extended objects 
whose separation is set by the 
correlation length of the field fluctuations. We therefore need a systematic procedure to eliminate 
the spurious zeros from the result in~\eqref{finalxnz}.  
One way is to restrict
the sums in~\eqref{alpha}, \eqref{beta} to those modes $c_n(t)$ that are {\it not} oscillating~\cite{Karra:1997he}, in other 
words to indices $n$ verifying
\be
\omega^{(n)}_2(t)\equiv\frac{4}{a^2}\sin^2 \left (\frac{\pi n}{N} \right ) + m_2(t) \leq 0\,.
\ee
It is indeed the presence of such unstable modes that allows for the production of the non-perturbative 
kink and anti-kink solutions. 
Then the formula for the average number density of kinks, $n_K$, is obtained by restricting the modes
that enter~\eqref{finalxnz}, giving us
\be
n_K= \frac{N}{2L}\left[1-\frac{2}{\pi} \sin^{-1}\left(\frac{\bar{\beta}}{\bar{\alpha}}\right)\right]\,,
\label{finalnk}
\ee
where now
\ba
\bar{\alpha} &\equiv& \sum_{\omega_2^{(n)}\leq 0} |c_n|^2 ,\label{alphabar} \\
\bar{\beta} &\equiv& \sum_{\omega_2^{(n)}\leq 0} |c_n|^2 \cos (2\pi n/N)\,.\label{betabar}
\ea
These equations only apply for $t \ge 0$ when the modes start to become unstable.
For $t < 0$, there are only fluctuating modes and we set 
$n_K=0$. We will discuss the difference 
between $\langle\hat{n}_Z\rangle$ and 
$n_K$ in Sec.~\ref{numerics}.

After the phase transition and as long as the lattice spacing $a$ is small enough, $a< 2/\sqrt{|m_2(t)|}$ 
for all times $t>0$, we can introduce $n_c(t)$, the time-dependent critical value of $n$ that separates 
unstable modes from modes that oscillate,
\be
n_c(t)\equiv \left\lfloor\frac{N}{\pi}\sin^{-1}\left(\frac{a\sqrt{|m_2(t)|}}{2}\right)\right\rfloor\,.
\label{ncrit}
\ee
where $\lfloor\,\rfloor$ denotes the integer part function.
Then $n_c(t)<N/2$ and~\eqref{alphabar}, \eqref{betabar} can be rewritten in a more explicit way 
\ba
\bar{\alpha} &\equiv& \sum_{|n|\leq n_c(t)} |c_n|^2 = |c_0|^2 + 2 \sum_{n=1}^{n_c(t)} |c_n|^2 ,
\label{baralpha} \\
\bar{\beta} &\equiv&\sum_{|n|\leq n_c(t)} |c_n|^2 \cos (2\pi n/N)\nn\\
&=& |c_0|^2+ 2\sum_{n=1}^{n_c(t)} |c_n|^2 \cos (2\pi n/N)\,. 
\label{barbeta}
\ea
Here we have identified $c_{-n}$ with $c_{N-n}$ for concision, and exploited the symmetry 
$c_{N-n}=c_n$ (valid for $1\leq n\leq N-1$) which can be checked directly via~\eqref{ckeq}, 
\eqref{ckt0}, \eqref{dotckt0}. Since the ratio $\bar{\beta}/\bar{\alpha}$ belongs to the interval $[0,1]$ one can also 
rewrite~\eqref{finalnk} as
\be
n_K= \frac{N}{\pi L}\cos^{-1}\left(\frac{\bar{\beta}}{\bar{\alpha}}\right)\,.
\label{finalnkbis}
\ee

Before diving into analytical and numerical estimates of 
$n_K$ we need to discuss the continuum and infinite volume limits of our discretized theory. We start with the continuum limit. Keeping $L$ fixed, and noticing that, for all $N$,
$n_c(t)\leq mL/4$, we can safely take the $N\rightarrow\infty$ limit in expressions involving $n/N$. In particular
\be
\omega^{(n)}_2(t)\approx\left(\frac{2\pi n}{L}\right )^2 + m_2(t) \,,
\ee
and 
\be
n_c(t)\approx\left\lfloor \frac{L\sqrt{|m_2(t)|}}{2\pi}\right\rfloor\,.
\ee
Then the expression for the ratio $\bar{\beta}/\bar{\alpha}$ reads
\be
\frac{\bar{\beta}}{\bar{\alpha}}\approx 
1-\frac{2\pi^2}{N^2}\frac{\sum_{\omega_2^{(n)}(t)\leq 0} n^2|c_n|^2}{\sum_{\omega_2^{(n)}(t)\leq 0} |c_n|^2}\,.
\ee
Now, it is clear that this expression is of the form $1-2x^2$ with $x\in[0,1]$ and therefore we may use the identity 
\be
\cos^{-1}\left(1-2x^2\right)=2\sin^{-1} x\,,
\ee
to simplify~\eqref{finalnkbis} and obtain
\ba
n_K &=& 
\frac{2N}{\pi L}\sin^{-1}\left(\frac{\pi}{N}
\sqrt{\frac{\sum_{\omega_2^{(n)}(t)\leq 0} n^2|c_n|^2}{\sum_{\omega_2^{(n)}(t)\leq 0} |c_n|^2}}\right)\nn\\
&=& \frac{2}{L}
\sqrt{\frac{\sum_{\omega_2^{(n)}(t)\leq 0} n^2|c_n|^2}{\sum_{\omega_2^{(n)}(t)\leq 0} |c_n|^2}}\,.
\ea
This is the expression of the continuum limit ($N\rightarrow\infty$) average number density of kinks. 
The main property of this expression is that is does not depend on $N$ anymore. Indeed, although 
the system's dynamics is governed by an infinite number of mode functions, only a finite number 
appears in the formula; it is only those modes with $n\leq mL/2\pi$ that trigger the instabilities 
required for the production of kinks. This means that the result is stable in the UV limit and does 
not depend on the resolution of our discretization. Physically, the contribution of the vacuum 
fluctuations of the quantum field has been discarded. 

Let us now end this section by discussing the infinite volume (or length since we are working in one 
spatial dimension) limit $L\rightarrow\infty$. This is readily done by noticing that the finite size of the 
spatial dimension is responsible for the discreteness of the wave vectors
\be
k_n\equiv\frac{2\pi n}{L}\,,
\ee
corresponding to different modes. As $L$ increases, however, these wave vectors become more and 
more numerous and densely packed until they form a continuum spanning the entire interval $[-m,m]$. 
At this point, it is convenient to switch notations and index any relevant quantities by $k_n$ instead of 
just $n$. Then
\be
\omega^{(k_n)}_2(t)= k_n^2 + m_2(t) \,,
\ee
and 
\be
k_c(t)\equiv \frac{2\pi n_c(t)}{L} = \sqrt{|m_2(t)|}\,.
\ee
The average kink number density can therefore be rewritten
\be
n_K=  \frac{1}{\pi}
\sqrt{\frac{\sum_{|k_n|\leq k_c(t)} k_n^2|c_{k_n}|^2}{\sum_{|k_n|\leq k_c(t)} |c_{k_n}|^2}}\,.
\label{preanalytical}
\ee
In the $L\rightarrow\infty$ limit the sums over $k_n$ become integrals over $k$, so that
\be
n_K \approx  
\frac{1}{\pi}\sqrt{\frac{\int_{0}^{k_c(t)} dk\,k^2|c_{k}|^2}{\int_{0}^{k_c(t)} dk\,|c_{k}|^2}}\,.
\label{analytical}
\ee
Here we have tacitly introduced the infinite volume mode functions $c_k$ verifying
\be
\ddot{c}_k + \left(k^2+m_2(t)\right)c_k=0\,,
\ee
and used their $k\rightarrow -k$ symmetry properties. 
We now have all the tools required to perform simple analytical estimates of the average kink 
number density.

\subsection{Analytical estimate}
\label{subsec: kinks_anal_est}

In the limit of a sudden phase transition ($\tau = 0$), we can solve~\eqref{ckeq} exactly 
since $m_2(t)=-m^2\Theta(t)$ (where $\Theta$ is the Heaviside function). In fact one may 
then choose the initial time to be $t_0=0_{-}$ and solve the differential equations
\be
{\ddot c}_n + \left [ \frac{4}{a^2}\sin^2 \left (\frac{\pi n}{N} \right ) - m^2 \right ] c_n = 0\,,
\label{ckeqsudden}
\ee
with initial conditions
\ba
c_n (0) &=& 
\frac{-i}{\sqrt{2L}} 
\left [ \frac{4}{a^2}\sin^2 \left ( \frac{\pi n}{N}\right ) + m^2 \right ]^{-1/4}\,,
\label{ckt0sudden}\\
{\dot c}_n (0) &=& \frac{1}{\sqrt{2L}} 
\left [ \frac{4}{a^2}\sin^2 \left (\frac{\pi n}{N} \right ) + m^2 \right ]^{1/4}\,.
\label{dotckt0sudden}
\ea
Since the time dependence of the frequency has disappeared, the above differential equations 
can be solved analytically. This yields the unstable mode functions $c_n(t)$ involved in the formula 
for the average number density of kinks~\eqref{finalnk} i.e. those verifying 
$|n|\leq N\sin^{-1}(ma/2)/\pi$. More precisely we have
\ba
c_n(t)&=&\frac{-i}{\sqrt{2L}}\left [ \frac{4}{a^2}\sin^2 
\left (\frac{\pi n}{N} \right ) + m^2 \right ]^{-1/4}\cosh\left(\kappa_n t\right)\nn\\
&&\hspace{-5mm}+\frac{1}{\sqrt{2L}}\left [ \frac{4}{a^2}\sin^2 \left (\frac{\pi n}{N} \right ) + m^2 \right ]^{1/4} 
\frac{\sinh(\kappa_n t)}{\kappa_n},
\ea
where $\kappa_n=\sqrt{m^2-\frac{4}{a^2}\sin^2 \left (\frac{\pi n}{N} \right)}$. 
Taking first the continuum limit $N\rightarrow\infty$ while keeping $L$ fixed we obtain, for $|n|\leq mL/2\pi$,
\ba
c_{k_n}(t)&\approx&\frac{-i}{\sqrt{2L}}\left (k_n^2 + m^2 \right)^{-1/4}\cosh\left( t \sqrt{m^2-k_n^2}\right)\nn\\
&&\hspace{-5mm}+\frac{1}{\sqrt{2L}}\left( k_n^2 + m^2 \right)^{1/4} 
\frac{\sinh\left( t\sqrt{m^2-k_n^2}\right)}{\sqrt{m^2-k_n^2}},
\ea
where we labelled the mode functions by $k_n=2\pi n/L$ as in the previous section. 
In the $L\rightarrow\infty$ limit, the discrete variable $k_n$ becomes continuous and we can write an 
analytical formula for the average kink number density as in~\eqref{analytical}:
\begin{widetext}
\be
n_K =  \frac{1}{\pi}\left\{\bigintsss_{0}^{m} dk\,
\left[\frac{k^2\left(m^2\cosh\left(2t\sqrt{m^2-k^2}\right)-k^2\right)}{(m^2-k^2)\sqrt{k^2+m^2}}\right]\right\}^{1/2}
\left\{\bigintsss_{0}^{m} dk\,
\left[\frac{m^2\cosh\left(2t\sqrt{m^2-k^2}\right)-k^2}{(m^2-k^2)\sqrt{k^2+m^2}}\right]\right\}^{-1/2}\,.
\label{analyticalsudden}
\ee
\end{widetext}
With this expression in hand we can immediately estimate a few important quantities. 
First of all, we can predict the late time behavior of the average kink number density. 
Indeed for large $t$ the integrals simplify considerably and it is easy to see that they are 
dominated by values of $k\ll m$. We can then 
estimate~\eqref{analyticalsudden} to be
\be
\label{kinks_analres}
n_K \approx  \frac{1}{\pi}\left[\frac{\int_{0}^{m} dk\,k^2\exp\left(-tk^2/m\right)}
{\int_{0}^{m} dk\,\exp\left(-tk^2/m\right)}\right]^{1/2}\approx \frac{1}{\pi}\sqrt{\frac{m}{2t}}
\ee
Using Eq.~\eqref{analyticalsudden}, one can also estimate the maximum number density of kinks 
that are produced after the phase transition. In fact, taking a time derivative of \eqref{analyticalsudden}, it is easy to convince oneself that this 
maximum occurs at $t=0_{+}$, in other words, immediately after the phase transition. 
Moreover its value can be computed exactly to be
\ba
\label{maxnkanal}
 n_K(0) &=&\frac{1}{\pi}\left(\frac{\int_{0}^{m} dk\, k^2/\sqrt{k^2+m^2}}{\int_{0}^{m}dk/\sqrt{k^2+m^2}}\right)^{1/2}\nn\\
&=& \frac{m}{\pi}\left(\frac{\sqrt{2}-\sinh^{-1}(1)}{2\coth^{-1}\left(\sqrt{2}\right)}\right)^{1/2}\approx 0.175 m\,.
\ea

Both the power law for the asymptotic behavior of the average kink number density and the maximum 
number of kinks value will be numerically confirmed in the following subsection. Our analytic results agree with previous work on sudden phase transitions in thermal quenches studied in~\cite{PhysRevE.48.767,Boyanovsky:1999wd,Ibaceta:1998yy,calzetta1989spinodal} using different techniques. 


\subsection{Numerical results}
\label{numerics}

We now discuss our numerical results for the time evolution of 
$n_K$ for different values of the quench parameter $\tau$. In principle this involves solving the complex 
differential equations \eqref{ckeq} with initial conditions \eqref{ckt0} and \eqref{dotckt0}, for the 
unstable mode functions $c_{n}(t)$ -- those with $|n|\leq n_c(t)$. We can then directly evaluate 
the average number density of kinks using \eqref{finalnk}. However, since this formula only involves 
$|c_n(t)|=\rho_n(t)$, considerable computational gain can be achieved by instead solving the real 
differential equations \eqref{rhok} with initial conditions \eqref{rhokic}, \eqref{rhokdotic}. 

It turns out that this system of ordinary differential equations presents a major computational 
difficulty caused by the fact that $\rho_{n}(t)$ grows exponentially for $|n|\leq n_c(t)$. Therefore 
the numerical evolution is limited to short time periods after the phase transition beyond which 
the numbers involved become extremely large and results cannot be trusted. One way to get 
around this problem is to factor out the exponential growth, {\it i.e.} the zero mode 
$\rho_0(t)=\rho_N(t)$, from the other modes and evolve it separately. So we write,
\be
\rho_{n}(t)=\rho_{0}(t)r_{n}(t)\,, 
\ee
for $n=1,\dots,N-1$. With this redefinition it can be shown that the differential equation (\ref{rhok}) now becomes,
\be
\label{req1}
\ddot{r}_{n}+2\frac{\dot{\rho}_0}{\rho_{0}}\dot{r}_{n}+\left( \omega_2^{(n)} - \omega_{2}^{(0)} 
+ \frac{1}{4 L^{2} \rho_{0}^{4}}\left( 1-\frac{1}{r_{n}^{4}} \right) \right) r_{n} = 0\,,
\ee
and its corresponding initial conditions are given by,
\ba
r_{n}(t_0) &=& \frac{1}{\sqrt{2L}} \frac{\omega_{2}^{(n)}(t_0)^{-1/4}}{\rho_{0}(t_0)}\,,\\
\dot{r}_{n}(t_0)&=&0\,.
\ea
Recall here that
\be
\label{omegan}
\omega_{2}^{(n)}=\frac{4}{a^{2}} \sin^{2}\left( \frac{\pi n}{N}\right) + m_{2}(t)\,,
\ee
and $\omega_2^{(0)}=\omega_{2}^{(N)}$.
Furthermore, one can also efficiently solve for $\rho_{0}(t)$ by introducing the auxiliary 
function $q(t) = \ln{\rho_0(t)}$, verifying
\be
\label{qeq}
\ddot{q}+\dot{q}^{2}+\omega_{2}^{(0)}=\frac{e^{-4q}}{4L^{2}}\,,
\ee
with initial conditions,
\ba
\label{qeqic}
q(t_0)&=&\ln\left[ \frac{1}{\sqrt{2L}}\Big( m_2 (t_0) \Big)^{-1/4}\right]\,,\notag\\
\dot{q}(t_0)&=&0\,.
\ea
By going to the $q(t)$ variable we avoid the exponential growth of $\rho_0(t)$.
Thus, both the differential equation for $r_{n}(t)$ \eqref{req1} and its corresponding initial conditions 
can be rewritten in terms of this auxiliary function:
\be
\label{reqmain}
\ddot{r}_{n}+2\dot{q}\dot{r}_{n}+\left(  \omega_2^{(n)} - \omega_{2}^{(0)}
+ \frac{ e^{-4q}}{4 L^{2}}\left( 1-\frac{1}{r_{n}^{4}} \right) \right) r_{n} = 0\,,
\ee
with initial conditions,
\ba
\label{reqmainic}
r_{n}(t_0) &=& \left(\frac{\omega_{2}^{(0)}(t_0)}{\omega_{2}^{(n)}(t_0)}\right)^{1/4}\,,\notag\\
\dot{r}_{n}(t_0)&=&0\,.
\ea
In summary, the numerically efficient way to study the dynamics of kink formation in our model, 
is to solve \eqref{qeq} and \eqref{reqmain} with respective initial conditions \eqref{qeqic} and 
\eqref{reqmainic}. The computational problem we had is indeed resolved since we managed to 
eliminate the exponential growth of $\rho_n(t)$ by suitable function redefinitions. The numerics 
can now be trusted for much longer periods of time.

 \begin{figure}[ht]
\includegraphics[width=0.45\textwidth,angle=0]{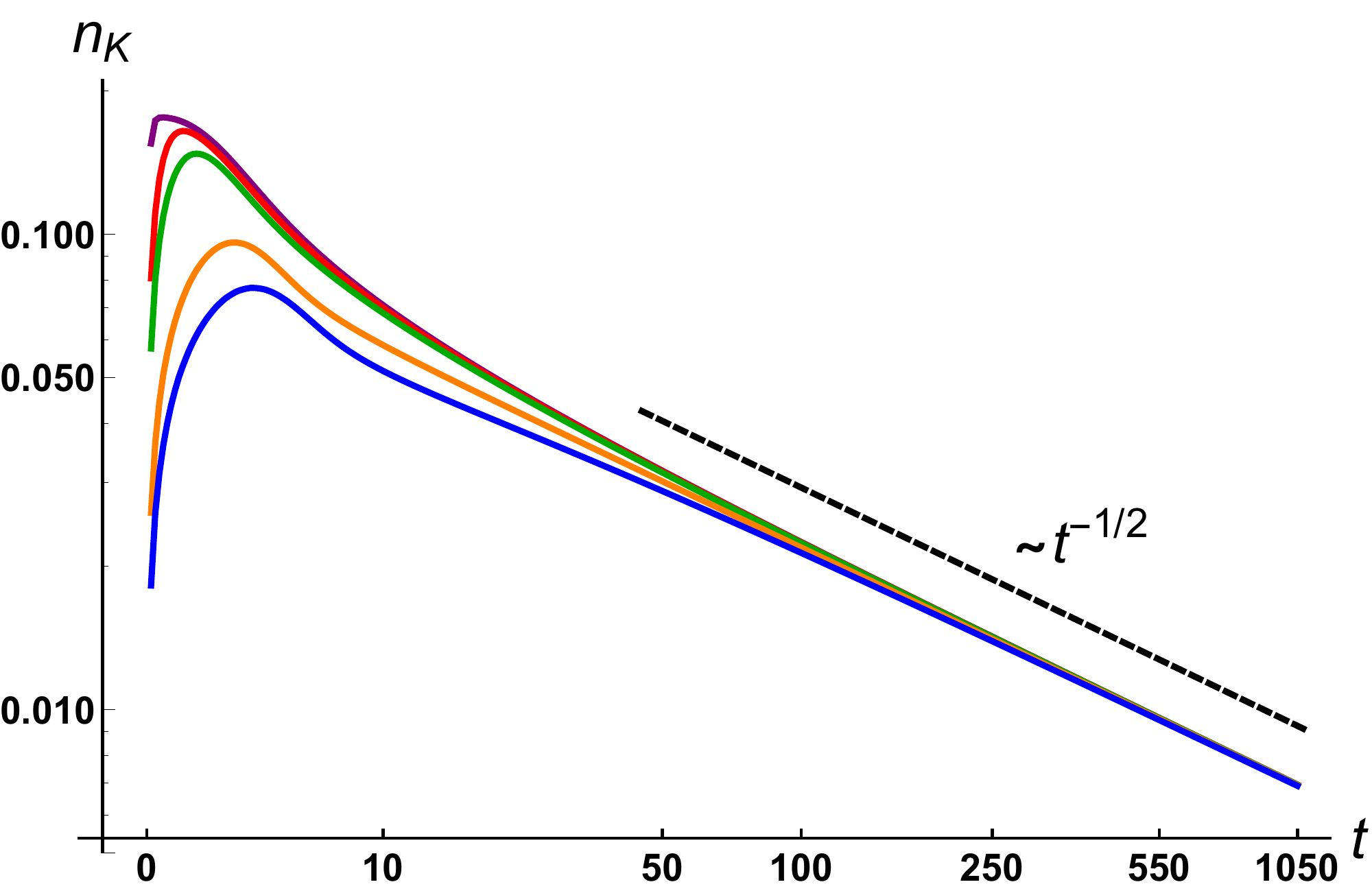}
\caption{Log-log plot of $n_K$ versus time for 
        $\tau$ =0.1 (Purple, topmost curve), 0.5 (Red), 1.0 (Green), 5.0 (Orange), 10.0 (Blue). The black dashed line shows the exhibited power law at late times, \emph{i.e.} $t^{-1/2}$ .}
\label{nKvstloglog}
\end{figure}

\begin{figure}[ht]
\includegraphics[width=0.45\textwidth,angle=0]{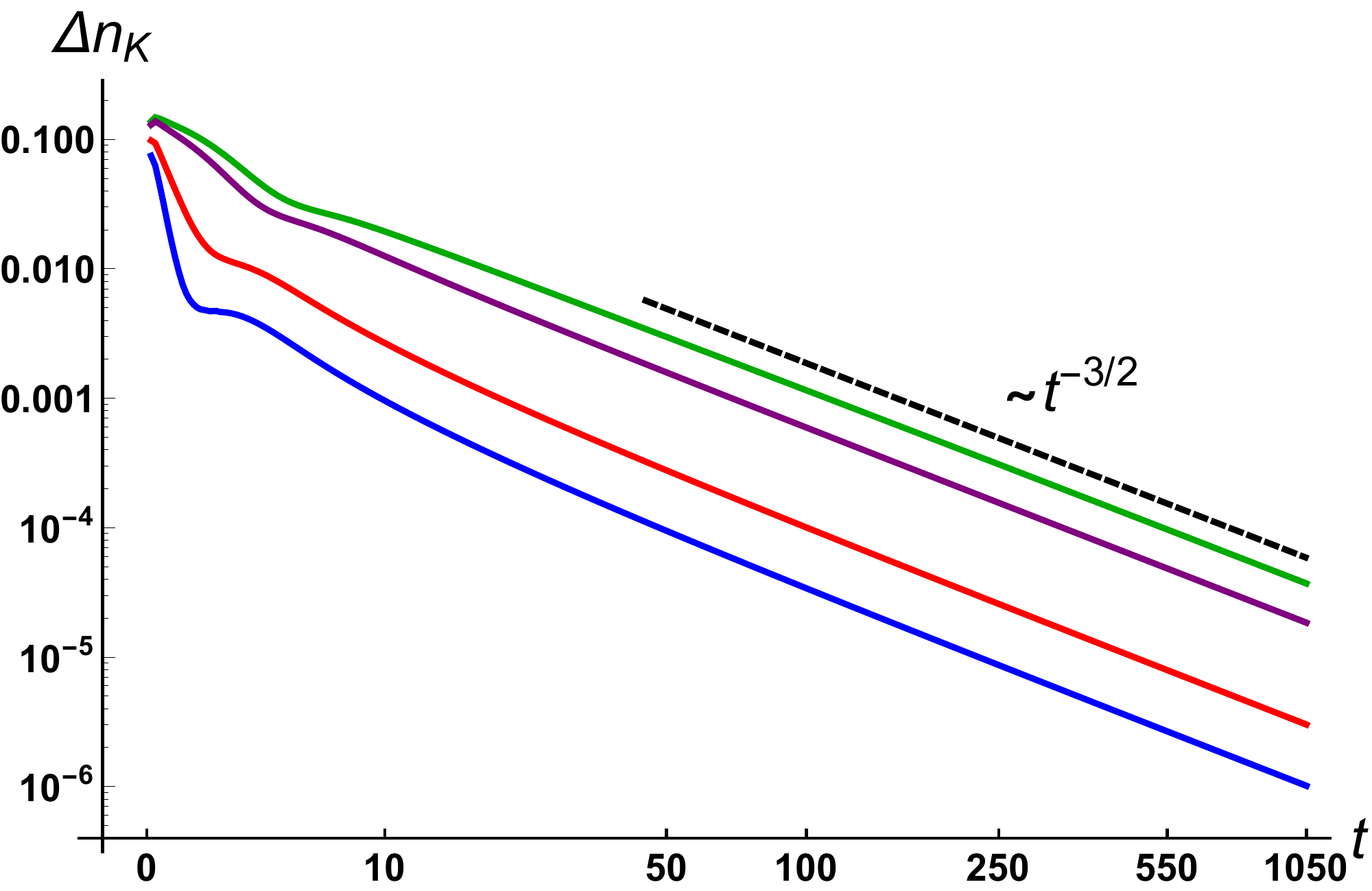}
\caption{Log-log plot of the differences between the average kink number density for different values of $\tau$, $n_K(t,\tau_1=0.1) - n_K(t,\tau_2)$ vs. time for $\tau_2 = $ 0.5 (Blue), 1.0 (Red), 5.0 (Purple), 10.0 (Green). The black dashed  line shows the exhibited power law, \emph{i.e.} $t^{-3/2}$. }
\label{difference}
\end{figure}

In our numerical work we work in units where $m=1$ and pick $t_0=-200$. To get accurate results we choose large $L$ and $N$. Most of our results are for $L=6400$ and $N=12800$. 
The evolution of the average number density of kinks  
$n_K$ for different quench time scales $\tau$ is shown in Fig.~\ref{nKvstloglog}.
The different curves exhibit the same qualitative behavior: immediately after the phase transition 
($t=0$) the average number density of kinks increases from $0$ to a maximum value 
$(n_{K})_\rmmax$
within a time $t_\rmmax$, and this is followed by a gradual decrease that asymptotically converges to 
a power law. Physically this corresponds to the production of a random distribution of kinks and anti-kinks 
during the phase transition, followed by their mutual annihilation over time. Noticeably, the asymptotic 
behavior of the average kink number density is independent of the quench time scale: at late times the 
plots for different values of $\tau$ converge to the {\it same}
function that falls off as
$t^{-1/2}$. (We have also cross-checked this result by computing the correlation length  
$\xi(t)$ of field fluctuations and showing that it scales as $1/n_K\sim t^{1/2}$, as expected from 
existing results in the literature~\cite{liu1991nonequilibrium,liu1992defect,PhysRevE.48.767,
Boyanovsky:1999wd,liu1992growth}. 
%
This scaling law also agrees with the analytical estimate of Eq.~\eqref{kinks_analres} and shows that the $\tau=0$ solution is a universal attractor.
To analyze the rate at which the kink densities for different values of $\tau$ converge,
we plot 
$\Delta n_K(t,\tau_1,\tau_2) \equiv n_K(t,\tau_1) - n_K(t,\tau_2)$ versus $t$ in Fig.~\ref{difference}.
We observe that at late times these differences fall off as $t^{-3/2}$. We can therefore conclude that 
\be
n_K (t) = C_K \sqrt{\frac{m}{t}} \, + \mathcal{O}\bigg(t^{-3/2}\bigg)\,,
\label{kinks_mainres}
\ee
where $C_K \approx 0.22$ is a constant of proportionality which is independent of the quench time 
scale $\tau$. This agrees well with the analytical estimate found in 
Eq.~\eqref{kinks_analres}: $1/(\pi \sqrt{2})\approx 0.225$. 

\begin{figure}[ht]
\includegraphics[width=0.45\textwidth,angle=0]{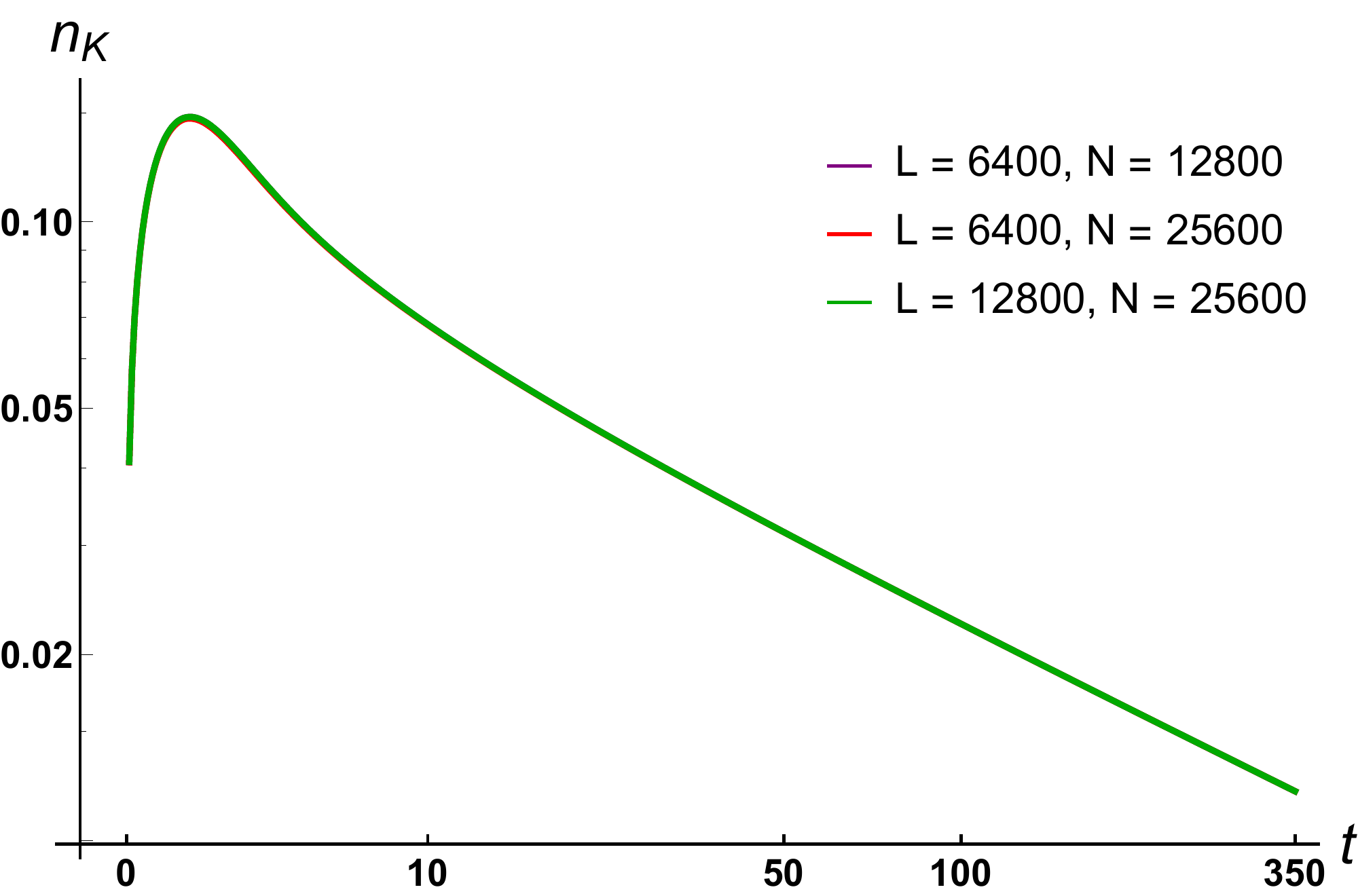}
\caption{Log-log plot of $n_{K} (t)$ versus time for $\tau =1.0$ for various values of $L$
and $N$ as given in \eqref{finalnk}.}
\label{cutoff1}
\end{figure}

\begin{figure}
\includegraphics[width=0.45\textwidth,angle=0]{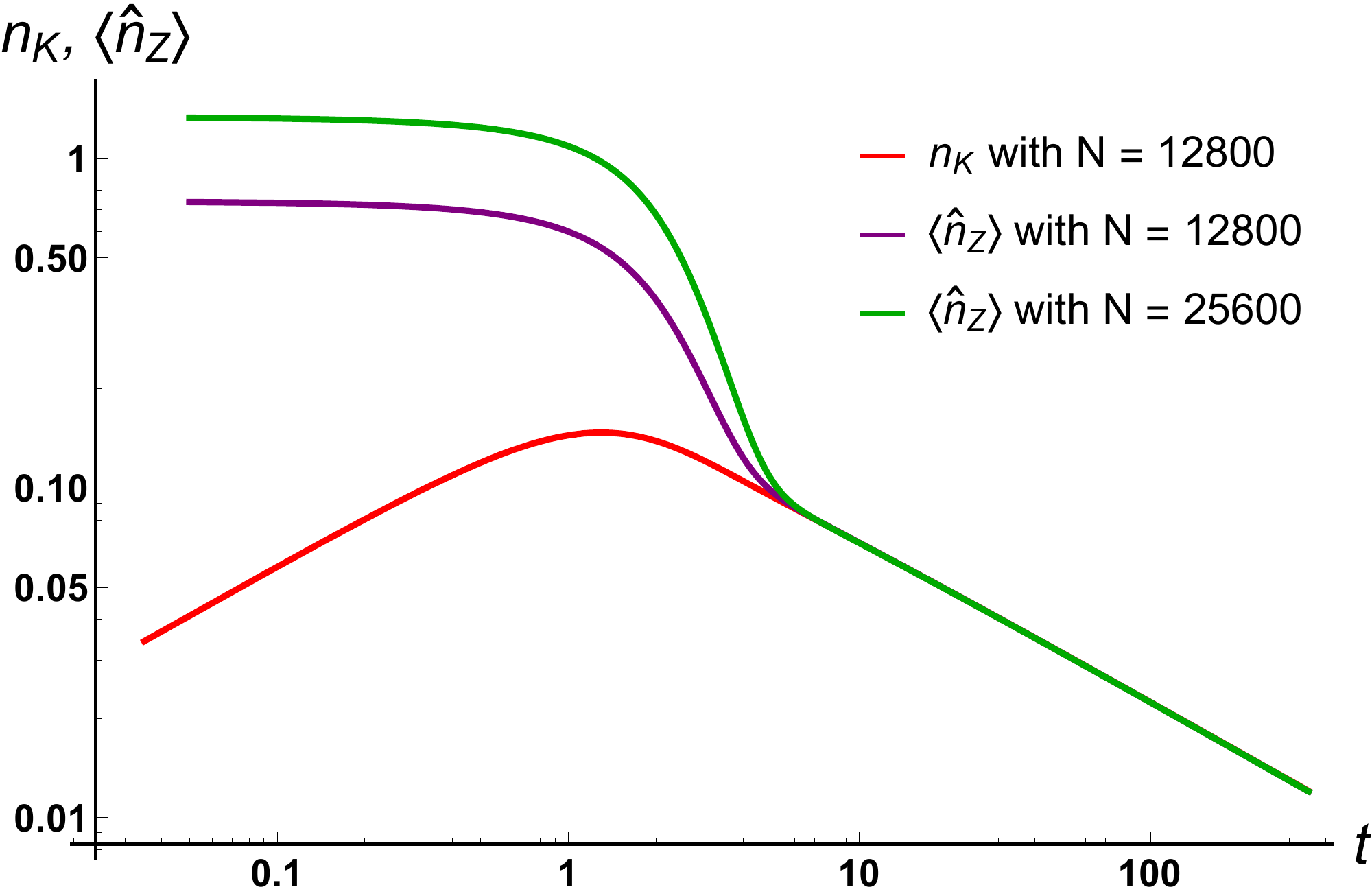}
\caption{Log-log plot of $n_{K}$ and $\langle\hat{n}_Z\rangle$ versus time for $\tau =1.0$, $L = 6400$ and various values of $N$.}
\label{cutoff2}
\end{figure}

We can explicitly check, as shown in Fig.~\ref{cutoff1}, that our results are independent 
 of both $L$ and $N$ as long as they are sufficiently large and $a=L/N$ is sufficiently small. 
 In Fig.~\ref{cutoff2} we have also plotted $\langle \hat{n}_Z \rangle$ and $n_{K}$ for different values of $N$. Although the curves depend on $N$ (or are UV sensitive) 
 near the phase transition, the late time behaviors are universal. This is to be expected since 
 unstable modes grow exponentially and dominate the sums in \eqref{baralpha} and
 \eqref{barbeta}. Thus our technique of restricting 
 the mode sums to differentiate between field zeros and kinks is reasonable and 
 gets rid of the artifacts arising due to finite $N$.
 
%
 
\begin{figure}[ht]
\includegraphics[width=0.45\textwidth,angle=0]{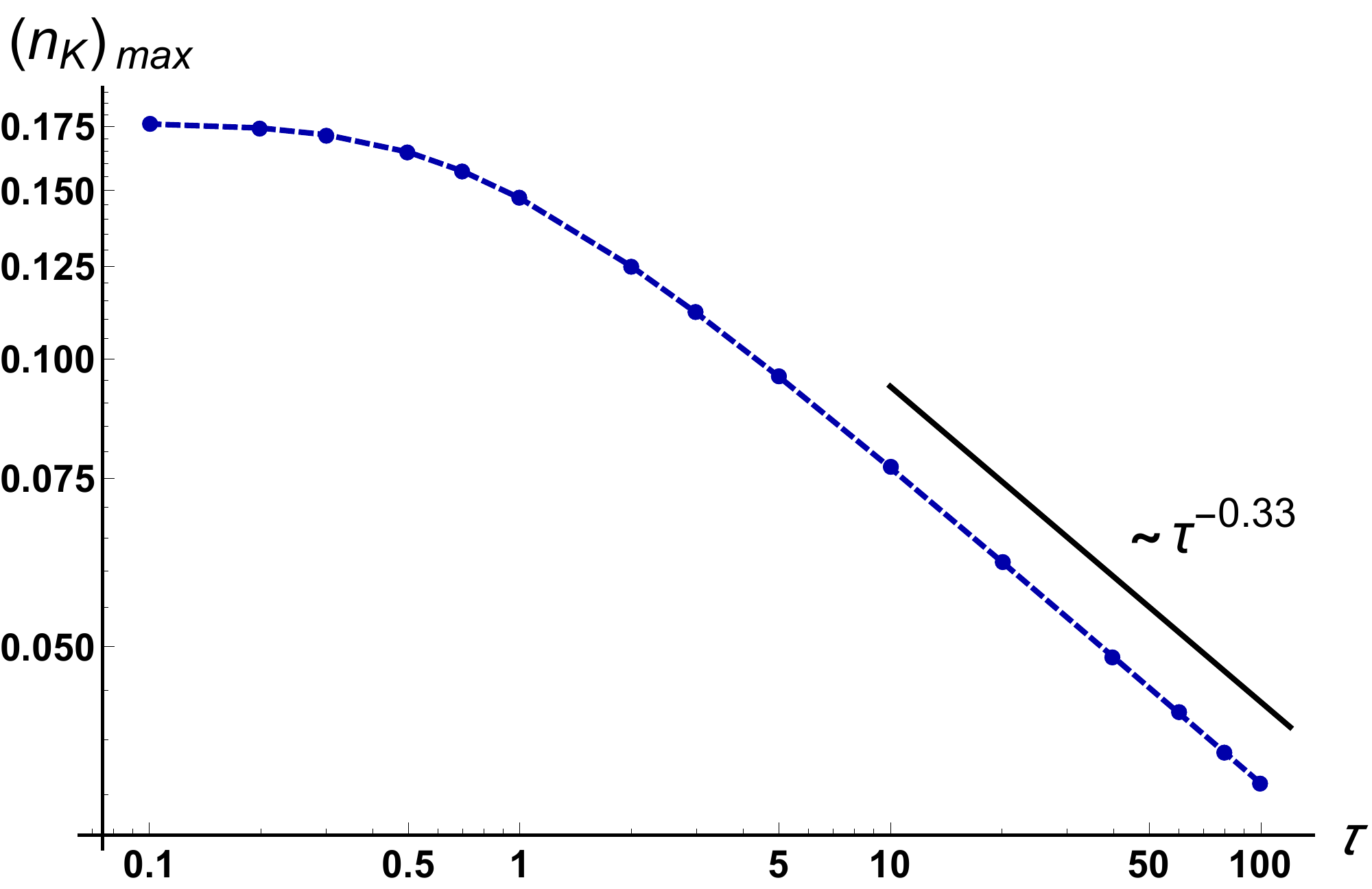}
\caption{Log-Log plot of the maximum average kink number density $(n_K)_{\rm max}$ vs. $\tau$. For larger values of $\tau$ the maximum average kink number density falls off as $\tau^{-0.33}$.}
\label{maxnkvstau}
\end{figure}
\begin{figure}
\includegraphics[width=0.45\textwidth,angle=0]{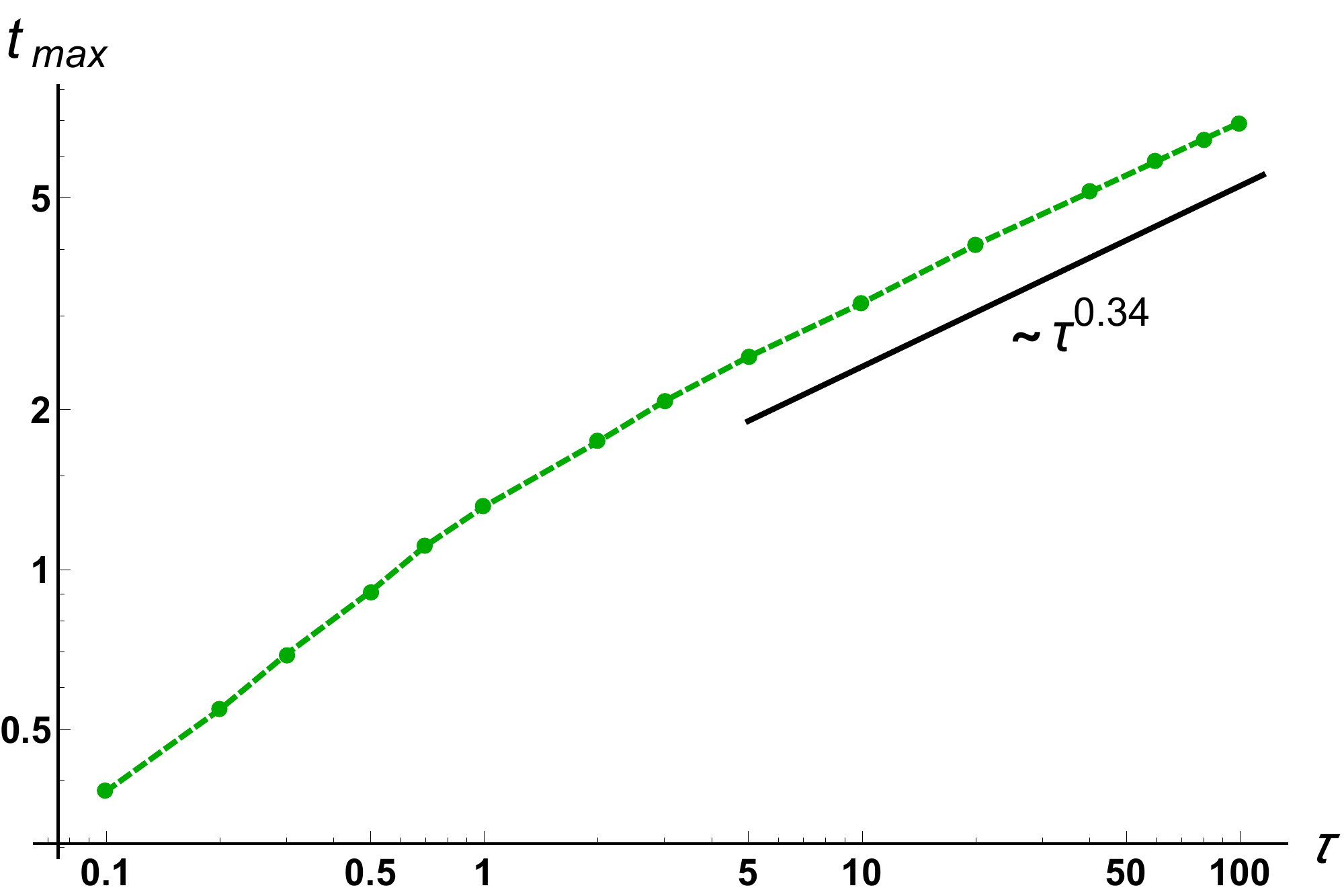}
\caption{Log-Log plot of the time at which maximum average kink number density $(n_K)_\rmmax$ occurs ($t_{\rm max}$) vs. $\tau$. For larger values of $\tau$, $t_\rmmax$ grows as $\tau^{0.34}$.}
\label{maxnkt}
\end{figure}

The plots of $(n_K)_\rmmax$ versus $\tau$, and $t_\rmmax$ 
versus $\tau$, are shown in Fig.~\ref{maxnkvstau} and Fig.~\ref{maxnkt} respectively. From these we note that the faster the 
phase transition (smaller quench time $\tau$), the more kinks and anti-kinks are produced 
and the faster their maximum number density is attained. In Fig.~\ref{maxnkvstau} we see 
that the maximum
density of kinks $(n_K)_\rmmax$ flattens, \emph{i.e.} it becomes a 
constant as quench time scales approach zero. The value of $(n_K)_\rmmax$ for which this happens 
is seen to be approximately $0.175$. This agrees remarkably well with the analytical estimate 
in Eq.~\eqref{maxnkanal}.

\section{Two Dimensions: Vortices}
\label{sec:vortices}

%

The analysis done in Section~\ref{kinks} can be generalized to the $d=2$ case. 
We will be considering a two-dimensional complex scalar field $\Phi$
whose dynamics are described by the Lagrangian density
\be
\label{vort_complex_lagrangian}
\mathcal{L}^{(2)}= \frac{1}{2}\partial_{\mu} \Phi^{*} \partial^{\mu} \Phi 
- \frac{1}{2}m_2^2 (t) \Phi^{*} \Phi -\frac{1}{4}\lambda \left(\Phi^*\Phi\right)^2\,.
\ee
This theory is known to possess solitonic solutions called {\it vortices}, characterized by a 
topological charge known as the {\it winding number}. Assuming a vortex field configuration 
$\Phi(x,y)=r(x,y) e^{i\theta(x,y)}=\phi(x,y)+i\psi(x,y)$ centered at a point $(x_0,y_0)$, 
 the winding number is given by
\be
\Gamma=\frac{1}{2\pi}\oint_{\cal C}d\theta = \frac{1}{2\pi}\int_{\cal C}\frac{1}{r^2}(\phi d\psi-\psi d\phi)\,,
\ee 
where $\mathcal{C}$ is any closed loop around $(x_0,y_0)$. Generically a non-zero winding number along 
a closed loop implies the existence of a vortex configuration and the vanishing of the field somewhere 
within the bounded region. Therefore, as in the case of kinks, vortices are to be found among 
zeros of  $\Phi$.

To study the production of vortices during the quantum phase transition we will thus do a 
similar analysis to the one we did for kinks. We start by setting $\lambda$ to zero and express 
the Lagrangian density in terms of the two real scalar fields $\phi$ and $\psi$, respectively defined 
as the real and imaginary part of the complex field $\Phi$:
\be
\label{vort_lagrangian}
\mathcal{L}^{(2)}= \half (\partial_\mu\phi)^2+\half (\partial_\mu\psi)^2 - \half m_2(t) (\phi^2+\psi^2)\,.
\ee
This is a model for two non-interacting real scalar fields in two spatial dimensions. In order to apply 
the methods outlined in Section~\ref{kinks}, we need to discretize this model. We first compactify both 
spatial dimensions by assuming periodic boundary conditions, $\phi(x+L,y)=\phi(x,y+L)=\phi(x,y)$ 
(and similarly for $\psi$). Space is thus seen to be a 2-torus of area $L^2$. We then discretize it on a 
regular square lattice consisting of $N^2$ points separated by a distance $a=L/N$ along both the $x$ 
and $y$ directions. Now for each lattice point $(x_{j},y_{l}) \equiv (ja,la)$ we can define the discretized 
fields  $\phi_{jl} \equiv \phi (x_{j},y_{l})$ and $\psi_{jl}\equiv  \psi (x_{j},y_{l})$. Writing the discretized 
Lagrangian and quantizing it can be done analogously to the one-dimensional case, with the 
understanding that any vectors and matrices are now $N^2$ and $N^2\times N^2$ dimensional 
respectively. For example, the vector of discretized field values of $\phi$ is given by 
\be
{\bm \phi} \equiv (\phi_{11},\phi_{12},...,\phi_{1N},
\phi_{21},...,\phi_{2N},...,\phi_{NN-1},\phi_{NN})^T.
\ee 
More generally, any $N^2\times N^2$ matrix $A$ will be represented by a two-dimensional array of 
matrix elements $A_{ij,kl}$ arranged in the following way:
\be
\label{fourindexmatrix}
A =
\begin{pmatrix}
A_{11,11}          & A_{11,12} & \cdots & A_{11,1N} & A_{11,21} & A_{11,22} & \cdots            \\
A_{12,11}          & A_{12,12} & \cdots & A_{12,1N} & A_{12,21} & A_{12,22} & \cdots             \\
\vdots  & \vdots  &    & \vdots  & \vdots   & \vdots    \\
A_{1N,11}          & A_{1N,12} & \cdots & A_{1N,1N} & A_{1N,21} & A_{1N,22} & \cdots             \\
A_{21,11}            &   &    &                           \\
A_{22,11}          &   &    &                               \\
  \vdots          &   &    &                                               
\end{pmatrix}
\nn
\ee
With these conventions in mind (where matrices are four index objects and vectors are two index 
objects), we can directly generalize the computations in Sec.~\ref{subsec:setup} to solve the 
functional Schr\"odinger equation for the wave-functional is $\Psi[\phi_{ij}, \psi_{ij};t]$. In fact, we 
can define a new $N^2\times N^2$ matrix $Z$ obeying Eqs.~\eqref{Zeq}, \eqref{Zic} and 
\eqref{Zdotic} as long as the matrix elements of $\Omega_2$ are given by
\be
[\Omega_2]_{ij,kl} = 
\begin{cases}
+{2}/{a^2}+m_2(t)\,,& i=k,j=l\\
-{1}/{a^2}\,,& i=k\pm1, j=l\pm1\ (\text{mod}\ N)\\
0\,,&\text{otherwise}\,.
\end{cases}
\ee
It is then easy to write the probability density functional as in Eq.~\eqref{probdensfin},
\be
|\Psi(t)|^2= \frac{1}{{\rm det}(2\pi K)} e^{- {\bm \phi}^T K^{-1} {\bm \phi} / 2}e^{- {\bm \psi}^T K^{-1} {\bm \psi} / 2}\,.
\label{probdensfin2}
\ee
where the matrix $K$ is still related to $Z$ via $K=ZZ^\dag$.

We can be even more explicit by realizing that the matrix $Z(t)$ is once again real and \emph{circulant}, 
{\it i.e.}, the matrix elements of $Z$, $Z_{pq,rs}$ depend only on $p-r$ (mod $N$) and $q-s$ (mod $N$). 
We can therefore again diagonalize $Z$ using the discrete Fourier transform:
\begin{widetext}
\be
\label{vort_zft}
Z_{pq,rs}=  \frac{1}{N} \sum_{n,n^{\prime}=1}^N c_{n,n^{\prime}}(t) e^{-i(p-r)2\pi n /N} e^{-i(q-s)2\pi n^{\prime} /N}\,.
\ee
Using equations (\ref{Zeq}),(\ref{Zic}) and (\ref{Zdotic}), the complex mode functions $c_{n,n^{\prime}}(t)$ verify
\be
\label{vort_ck}
{\ddot c}_{n,n^{\prime}} + \left [ \frac{4}{a^2} \left\{ \sin^2 \left (\frac{\pi n}{N} \right ) 
+ \sin^2 \left (\frac{\pi n^{\prime}}{N} \right ) \right\} + m_2(t) \right ] c_{n,n^{\prime}} = 0\,,
\ee
and 
\be
c_{n,n^\prime} (t_0) = 
\frac{-i}{\sqrt{2a}}\frac{1}{N} 
\left [ \frac{4}{a^2} \left\{ \sin^2 \left ( \frac{\pi n}{N}\right ) 
+ \sin^2 \left ( \frac{\pi n^{\prime}}{N}\right )  \right\} + m_2(t_0) \right ]^{-1/4}\,,
\label{vort_ic}
\ee
\be
{\dot c}_{n,n^{\prime}} (t_0) = \frac{1}{\sqrt{2a}} \frac{1}{N} 
\left [ \frac{4}{a^2} \left\{ \sin^2 \left (\frac{\pi n}{N} \right ) 
+  \sin^2 \left (\frac{\pi n^{\prime}}{N} \right ) \right\} + m_2(t_0) \right ]^{1/4}\,.
\label{vort_icdot}
\ee
\end{widetext}
Note that $c_{n,n'}=c_{n',n}$ which immediately implies that $Z_{pq,rs}=Z_{qp,sr}$ and again we 
assume the initial time $t_0$ to be such that $t_0\ll -\tau$. This follows from the rotational symmetry of the system.


\subsection{Average vortex number density}
\label{vort_avgdensity}


To find the vortex number density, we first need a quantum operator that counts the number 
of zeros $n_Z$ of the complex field $\Phi$ (as in Sec.~\ref{subsec:kinks_avg}), or in other 
words, coincident zeroes of both the fields $\phi$ and $\psi$. Since 
space is discretized, such an operator necessarily yields a coarse-grained estimate of the actual 
number of zeros of a given field configuration. 
As the number of lattice points 
$N^2$ increases so does the operator's resolution: while certain ``zeros'' cease to be counted, 
new ones are revealed. In the limit where $N\rightarrow \infty$ we expect divergences,
just as in the kink case, and we will return to this point later on.

\begin{figure}[t]
      \includegraphics[width=0.35\textwidth,angle=0]{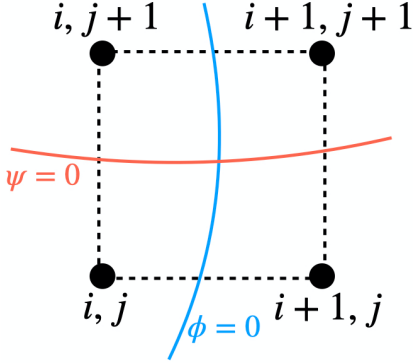}
  \caption{A plaquette showing how zeros are counted.}
  \label{fig:vort_plaquette}
\end{figure}

We think of the vortex as the intersection of a domain wall of $\phi$ -- for our purposes,
a domain wall is a curve on which $\phi=0$ --  with a domain wall
of $\psi$. Then, as shown in Fig.~\ref{fig:vort_plaquette}, there could be a situation where
a $\phi$ domain wall enters a plaquette through one edge and leaves through the opposite
edge, while a $\psi$ domain wall passes through the plaquette in the orthogonal direction.
Then the two domain walls must intersect, leading to coincident zeros that correspond to
a vortex within that plaquette. Other possibilities include the case where the $\phi$ wall
enters the plaquette from the lower edge but leaves from the right edge in Fig.~\ref{fig:vort_plaquette}
while the $\psi$ domain wall goes through as shown or bends to exit from the top edge.
It is ambiguous whether a coincident zero exists in these other cases but the ambiguity
is minimized as the lattice resolution is increased ($N \to \infty$). Hence we can count
zeros of $\Phi$ in the large $N$ limit by counting the plaquettes in which $\phi$ and $\psi$ 
domain walls enter across orthogonal edges.

Then, motivated by the discussion in Sec.~\ref{subsec:kinks_avg}, we can define the 
number density of  zeros of $\Phi$ by
\begin{widetext}
\ba
\label{vort_count_op}
\hat{n}_{Z} &=& n_{Z}\left(\hat{\bm\phi},\hat{\bm\psi}\right) 
\equiv \frac{1}{L^2} \sum_{i,j=1}^N \frac{1}{16} \biggl[
 \left\{  \text{sgn}\left(\hat{\phi}_{ij}\right) 
 - \text{sgn}\left(\hat{\phi}_{i+1,j}\right) \right\}^2 \left\{  \text{sgn}\left(\hat{\psi}_{ij}\right) 
 - \text{sgn}\left(\hat{\psi}_{i,j+1}\right) \right\}^2 +\ (\phi\leftrightarrow \psi)\  \biggr] \\
 &=& \frac{1}{4 L^2} \sum_{i,j=1}^N \biggl[
 \left\{  1 - \text{sgn}\left(\hat{\phi}_{i,j} \hat{\phi}_{i+1,j} \right) \right\} 
 \left\{ 1 - \text{sgn}\left(\hat{\psi}_{ij} \hat{\psi}_{i,j+1} \right) \right\} 
 + \left\{ 1 - \text{sgn}\left(\hat{\psi}_{ij}  \hat{\psi}_{i+1,j} \right) \right\} 
 \left\{ 1 - \text{sgn}\left(\hat{\phi}_{ij} \hat{\phi}_{i,j+1} \right) \right\}  \biggr]\nn\,.
\ea
\end{widetext}
%

We can now write down the vacuum expectation value of the operator $\hat{n}_{Z}$:
$\langle \hat{n}_{Z}\rangle \equiv \langle 0|\hat{n}_{Z}(t)|0\rangle$.
Using the fact that the fields $\phi$ and $\psi$ are independent and that, consequently, 
the probability density functional factorizes as in Eq.~\eqref{probdensfin2}, we first notice that
\ba
\left\langle\text{sgn}\left(\hat{\phi}_{ij}\hat{\phi}_{i,j+1}\right)\right\rangle &=& \frac{1}{\sqrt{{\rm det}(2\pi K)}}\times\nn\\
&&\hspace{-2cm}\int d^N{\bm \phi}\ \text{sgn}\left(\phi_{ij}\phi_{i,j+1}\right) e^{- {\bm \phi}^T K^{-1} {\bm \phi} / 2}\,.
\label{corrsgn}
\ea
Then, using the fact that the matrix $K^{-1}$ is circulant and, moreover, symmetric under interchange of its first (or last) two indices -- properties that are inherited from $Z$, we can establish that
\ba
\left\langle\text{sgn}\left(\hat{\phi}_{ij}\hat{\phi}_{i,j+1}\right)\right\rangle&=&\left\langle\text{sgn}\left(\hat{\phi}_{11}\hat{\phi}_{12}\right)\right\rangle\nn\\
&=&\left\langle\text{sgn}\left(\hat{\phi}_{11}\hat{\phi}_{21}\right)\right\rangle\nn\\
&=&\left\langle\text{sgn}\left(\hat{\phi}_{ij}\hat{\phi}_{i+1,j}\right)\right\rangle\,.
\ea
Physically, this set of equalities is a manifestation of the translational and rotational invariance of the system. It is also clear that, $\phi$ being a dummy variable in the integral of Eq.~\eqref{corrsgn},
\be
\left\langle\text{sgn}\left(\hat{\psi}_{ij}\hat{\psi}_{kl}\right)\right\rangle = \left\langle\text{sgn}\left(\hat{\phi}_{ij}\hat{\phi}_{kl}\right)\right\rangle\,.
\ee
%
These properties thus allow us to write the average number of zeros of the field in a very simple form:
\be
\label{vort_op_avg}
\langle \hat{n}_{Z} \rangle = \frac{N^2}{2 L^2} \biggl[
 1 - \left\langle \text{sgn}\left(\hat{\phi}_{11} \hat{\phi}_{12} \right) \right\rangle \biggr]^2\,.
\ee

From this point on, the computation of the average number of zeros follows along the same lines 
as in Sec.~\ref{kinks}, and we obtain
\be
\langle \hat{n}_{Z}\rangle = \frac{N^2}{2L^2}\left[1-\frac{2}{\pi} \sin^{-1}\left(\frac{\beta}{\alpha}\right)\right]^2\,,
\label{vort_finalnz}
\ee
where $\alpha$ and $\beta$ are now defined as
\ba
\alpha &\equiv& K_{11,11} = \sum_{n,n^{\prime}=1}^N |c_{n,n^{\prime}}|^2 ,\label{vort_alpha} \\
\beta &\equiv& K_{11,12} = \sum_{n,n^{\prime}=1}^N |c_{n,n^{\prime}}|^2 \cos (2\pi n^{\prime}/N)\,.
\label{vort_alphabeta}
\ea
Here, once again, we have used the reality of the matrix $Z$. 

Eq.~\eqref{vort_finalnz} gives us the number density of field zeros but
we are interested in counting the number density of {\it vortices}. 
We have already discussed how quantum fluctuations can induce a non-zero number density of zeros 
of the field even in the absence of spontaneous symmetry breaking. We thus need to eliminate such 
spurious zeros by restricting the sums in \eqref{vort_alpha} and (\ref{vort_alphabeta}) to the mode functions $c_{n,n^{\prime}}(t)$ 
that are \emph{non-oscillating}. In this case we include the modes corresponding to $n$ and $n^{\prime}$ 
verifying,
\be
\omega^{(n,n^{\prime})}_2(t)\equiv \frac{4}{a^2} \left\{ \sin^2 \left ( \frac{\pi n}{N}\right ) 
+ \sin^2 \left ( \frac{\pi n^{\prime}}{N}\right )  \right\} + m_2(t) \leq 0\,.
\ee
The average number density of vortices formed after the phase transition is finally given by,
\be
n_V = \frac{N^2}{2L^2}
\left[1-\frac{2}{\pi} \sin^{-1}\left(\frac{\bar{\beta}}{\bar{\alpha}}\right)\right]^2\,,
\label{vort_finalnv}
\ee
where,
\ba
\bar{\alpha} &\equiv& \sum_{\omega_2^{(n,n^{\prime})}\leq 0} |c_{n,n^{\prime}}|^2 ,\label{vort_alphabar} \\
\bar{\beta} &\equiv& \sum_{\omega_2^{(n,n^{\prime})}\leq 0} |c_{n,n^{\prime}}|^2 \cos (2\pi n^{\prime}/N)\,.
\label{vort_betabar}
\ea
Similar to the case of kinks (see discussion in Sec.~\ref{subsec:kinks_avg}), this 
result only makes sense after the phase transition; it is ill-defined before.
As might be intuitively expected, the average number density of vortices is obtained, up to a combinatorics factor due to the $\phi\leftrightarrow\psi$ symmetry, by squaring 
the average number density of kinks. In the next subsections we will see that this intuition is supported by 
both analytical and numerical estimates of the asymptotic dynamics of the problem.

Analogously to Sec.~\ref{subsec:kinks_avg}, Eq.~\eqref{vort_finalnv} can be further simplified by first 
taking the continuum limit $N\rightarrow\infty$ (at fixed volume $L$) to obtain
\be
n_V
\approx  \frac{8}{L^2}
\frac{\sum_{\omega_2^{(n,n')}\leq 0} n'^2|c_{n,n'}|^2}{\sum_{\omega_2^{(n,n')}\leq 0} |c_{n,n'}|^2}\,,
\label{preanalytic2}
\ee
where the sums run over pairs of integers $(n, n')\in \mathbb{Z}^2$ verifying
\be
\omega^{(n,n^{\prime})}_2(t)\approx \left(\frac{2\pi n}{L}\right)^2 + \left(\frac{2\pi n'}{L}\right)^2 + m_2(t_0) \leq 0\,,
\ee
and it is understood that $c_{-n,n'}\equiv c_{N-n,n'}$, $c_{n,-n'}\equiv c_{n,N-n'}$ for 
$0\leq n,n'\leq N-1$. Relabelling the mode functions by the discrete two-dimensional wave vector 
\be
\vec{k}_{n,n'} = \left( k_x^{(n)}, k_y^{(n')}\right) \equiv \left(\frac{2\pi n}{L}, \frac{2\pi n'}{L}\right)\,,
\ee
and taking the large $L$ limit, Eq.~\eqref{preanalytic2} can be recast as
\be
\label{geoabove}
n_V \approx  
\frac{2}{\pi^2}\frac{\int_{k\leq k_c(t)} d^2 k\, k_y^2|c_{\vec{k}}|^2}{\int_{k\leq k_c(t)} d^2 k\, |c_{\vec{k}}|^2}\,.
\ee
Here we have once again introduced the infinite volume mode functions 
$c_{\vec{k}}$ -- labelled by a continuum of two-dimensional wave vectors 
$\vec{k}\equiv (k_x,k_y)$ -- verifying
\be
\ddot{c}_{\vec{k}} + \left(k^2+m_2(t)\right)c_{\vec{k}}=0\,,
\ee
and defined $k\equiv|\vec{k}|$ and $k_c(t)=\sqrt{|m_2(t)|}$ as in Sec.~\ref{subsec:kinks_avg}. 
Noticing
that $c_{\vec{k}}$ only depends\footnote{This can be checked explicitly using Eqs.~\eqref{vort_ck}, 
\eqref{vort_ic}, \eqref{vort_icdot} and is a consequence of the rotational invariance of the problem.} 
on $k$ and going to polar coordinates, we can turn the double integrals in \eqref{geoabove} into single integrals
to finally obtain the continuum, infinite volume limit of the average vortex number density:
\be
n_V \approx  
\frac{1}{\pi^2}\frac{\int_{0}^{k_c(t)} dk\, k^3 |c_{\vec{k}}|^2}{\int_{0}^{k_c(t)} dk\, k|c_{\vec{k}}|^2}\,.
\label{analytic2}
\ee
With the possible exception of our particular choice of UV cutoff, this formula is in agreement 
with known results in the literature (see {\it e.g.} Eq.~(5) in \cite{Ibaceta:1998yy}).

\subsection{Analytical estimate}

\label{subsec: vort_anal_est}

Just as we did in the case of kinks in Sec.~\ref{subsec: kinks_anal_est}, we can also compute the average number density of vortices 
at late times in the limit of a sudden phase transition 
($\tau = 0$). This can be achieved once again by exactly solving the differential equations for the mode 
coefficients $c_{n,n^{\prime}}(t)$. As we saw in Sec.~\ref{sec:vortices} these differential equations 
are as follows, 
\be
{\ddot c}_{n,n^{\prime}} + \left [ \frac{4}{a^2} \left\{ \sin^2 \left ( \frac{\pi n}{N}\right ) 
+ \sin^2 \left ( \frac{\pi n^{\prime}}{N}\right )  \right\} - m^2 \right ] c_{n,n^{\prime}} = 0\,,
\label{vort_ckeqsudden}
\ee
with initial conditions
\begin{widetext}
\ba
c_{n,n^{\prime}} (0) &=& 
\frac{-i}{\sqrt{2a}} \frac{1}{N} 
\left [ \frac{4}{a^2} \left\{ \sin^2 \left ( \frac{\pi n}{N}\right ) 
+ \sin^2 \left ( \frac{\pi n^{\prime}}{N}\right )  \right\} + m^2 \right ]^{-1/4}\,,
\label{vort_ckt0sudden}\\
{\dot c}_{n,n^{\prime}} (0) &=& \frac{1}{\sqrt{2a}} \frac{1}{N}
\left [ \frac{4}{a^2} \left\{ \sin^2 \left ( \frac{\pi n}{N}\right ) 
+ \sin^2 \left ( \frac{\pi n^{\prime}}{N}\right )  \right\} + m^2 \right ]^{1/4}\,.
\label{vort_dotckt0sudden}
\ea
The solution to these equations can be obtained analytically. In fact they look very similar to the ones 
we obtained in the kinks case but now involve two indices instead of just one. This gives the unstable 
mode functions $c_{n,n^{\prime}}(t)$ involved in the formula for the average number density of vortices:
\ba
c_{n,n^{\prime}}(t)&=&\frac{-i}{\sqrt{2L}}\left [ \frac{4}{a^2} \left\{ \sin^2 \left ( \frac{\pi n}{N}\right ) 
+ \sin^2 \left ( \frac{\pi n^{\prime}}{N}\right )  \right\} 
+ m^2 \right ]^{-1/4}\cosh\left(\kappa_{n,n^{\prime}} t\right)\nn\\
&&+\frac{1}{\sqrt{2L}}\left [ \frac{4}{a^2} \left\{ \sin^2 \left ( \frac{\pi n}{N}\right ) 
+ \sin^2 \left ( \frac{\pi n^{\prime}}{N}\right )  \right\} + m^2 \right ]^{1/4} 
\frac{\sinh(\kappa_{n,n^{\prime}} t)}{\kappa_{n,n^{\prime}}},
\ea
where 
\be
\kappa_{n,n^{\prime}}=\sqrt{m^2-\frac{4}{a^2} \left\{ \sin^2 \left ( \frac{\pi n}{N}\right ) 
+ \sin^2 \left ( \frac{\pi n^{\prime}}{N}\right )  \right\} } .
\ee
Now, taking first the continuum limit 
$N\rightarrow\infty$ while keeping $L$ fixed we obtain, for $n^2 + n^{\prime 2}\leq (mL/2\pi)^2$ ,
\ba
c_{\vec{k}_{n,n'}}(t)&\approx&
\frac{-i}{\sqrt{2L}}\left (k_x^{(n)}{}^2 + k_y^{(n')}{}^2 + m^2 \right)^{-1/4}
\cosh\left( t \sqrt{m^2-k_x^{(n)}{}^2- k_y^{(n')}{}^2}\right)\nn\\
&&+\frac{1}{\sqrt{2L}}\left( k_x^{(n)}{}^2+ k_y^{(n')}{}^2 + m^2 \right)^{1/4} 
\frac{\sinh\left( t\sqrt{m^2-k_x^{(n)}{}^2 - k_y^{(n')}{}^2}\right)}{\sqrt{m^2-k_x^{(n)}{}^2 - k_y^{(n')}{}^2}},
\ea
where we have relabelled the mode functions by $\vec{k}_{n,n'}=(k_x^{(n)},k_y^{(n')})$ and recall that 
$k_x^{(n)}=2\pi n/L$, $ k_y^{(n')}=2\pi n^{\prime}/L$.
In the limit $L\rightarrow\infty$, the discrete variables $\vec{k}_{n,n'}$ become continuous and, as 
in Eq.~\eqref{analytic2}, we can write an analytical formula for the average number density of vortices:
\be
n_V \approx  
\frac{1}{\pi^2}\left\{\bigintsss_{0}^{m} dk \left[\frac{k^3\left(
m^2\cosh\left(2t\sqrt{m^2-k^2}\right)-k^2 \right)}{(m^2-k^2)\sqrt{k^2+m^2}}\right]\right\}
 \left\{\bigintsss_{0}^{m} dk  
 \left[\frac{k \left( m^2\cosh\left(2t\sqrt{m^2-k^2}\right)-k^2\right)}{(m^2-k^2)\sqrt{k^2+m^2}}\right]\right\}^{-1}\,.
 \label{vort_analyticalsudden}
\ee
\end{widetext}
Using this equation, we can once again estimate the late time behavior of the average number of 
vortices. In the limit, $k,k^{\prime} \ll m$, we have
\be
n_V \approx  \frac{1}{\pi^2}\frac{\int_{0}^{m} dk\,k^3\exp\left(-tk^2/m\right)}
{\int_{0}^{m} dk\,k\exp\left(-tk^2/m\right)} \approx \frac{m}{\pi^2t}=2!\,n_K^2\,.
\label{vortanalres}
\ee
As mentioned below Eq.~\eqref{vort_betabar}, the vortex number density is obtained by squaring the kink number density and multiplying by the combinatorial factor $2!$ due to the exchange symmetry $\phi\leftrightarrow\psi$.

Furthermore, like in the case of kinks, the maximum number density of vortices can be estimated 
using Eq.~\eqref{vort_analyticalsudden}. This maximum is reached immediately after the phase 
transition, at time $t=0_{+}$ and is found to be
\ba
(n_V)_\rmmax &=& \frac{1}{\pi^2} \frac{\int_0^m dk\, k^3/\sqrt{k^2+m^2} }
{\int_{0}^{m} dk\, k/\sqrt{k^2+m^2} } \nn \\
&=& \frac{m^2\sqrt{2}}{3\pi^2} \approx 0.0478\, m^{2}\,.
\label{maxnvanal}
\ea

\subsection{Numerical results}
\label{vort_num_res}

We use numerical techniques to solve (\ref{vort_ck}) and then calculate the average vortex 
number density using (\ref{vort_finalnv}). For reasons discussed earlier, the parameters $L$ 
and $N$ that we choose for our numerical simulations need to be sufficiently large to accurately 
describe the continuum infinite volume limit. We choose, $L=2000$ and $N=4000$. As in the 
case of kinks, the results are insensitive to the UV and IR cutoffs. In practice, because of the 
order $N^2$ computational complexity of the problem and the exponential growth of the magnitudes 
of mode functions, we directly solve for $\rho_{n,n'}=|c_{n,n'}|$
and factor out the zero mode to improve 
the numerical accuracy (see Sec.~\ref{numerics} for details).

\begin{figure}[ht]
  \includegraphics[width=0.45\textwidth,angle=0]{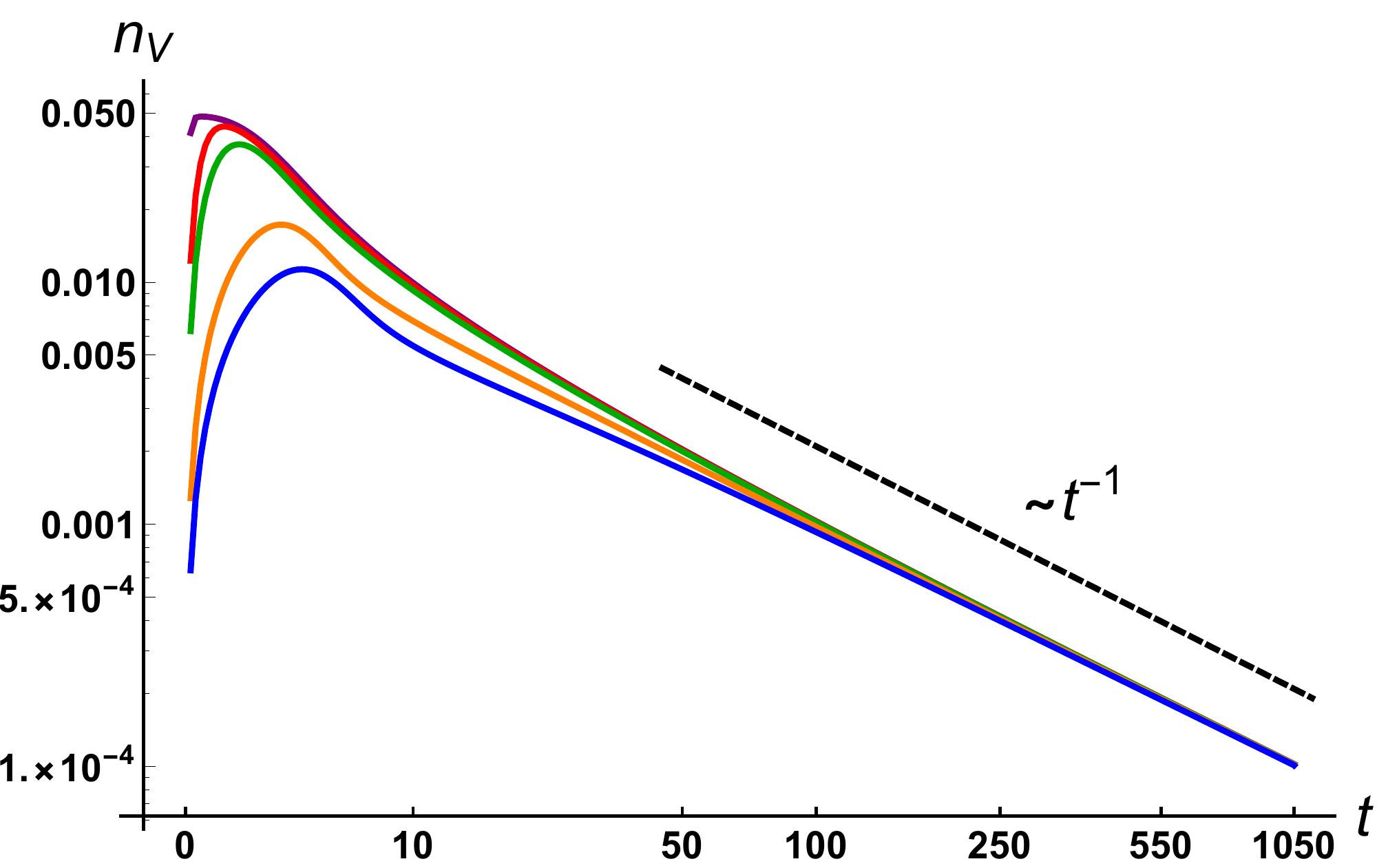}
\caption{Log-log plot of  $n_V(t)$ versus time for $\tau$ =0.1 (Purple, topmost curve), 0.5 (Red), 1.0 (Green), 5.0 (Orange), 10.0 (Blue). The black dashed line shows the exhibited power law at late times, \emph{i.e.} $t^{-1}$.}
\label{vort_nKvstloglog}
\end{figure}

\begin{figure}[ht]
   \includegraphics[width=0.45\textwidth,angle=0]{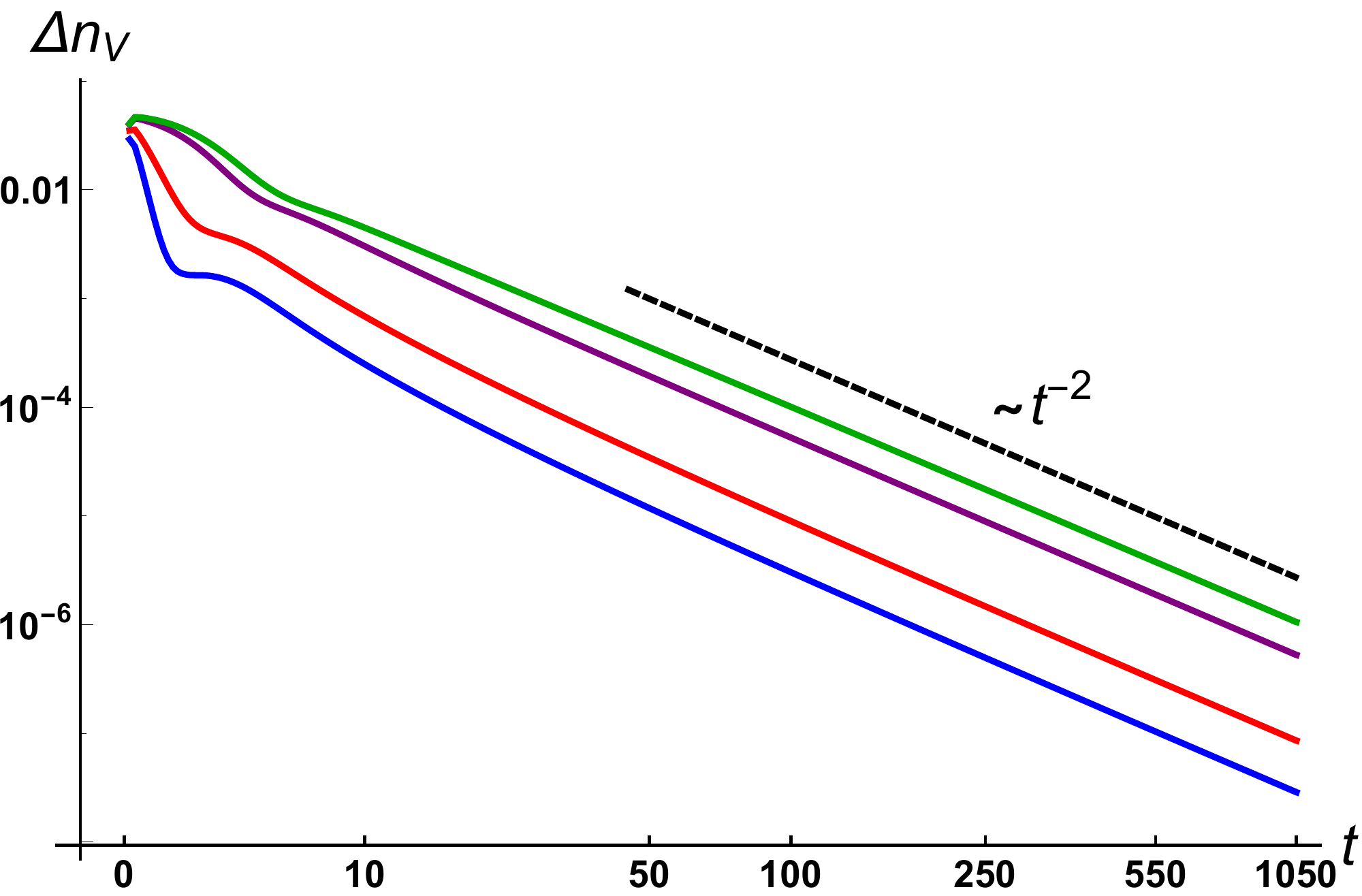}
  \caption{Log-log plot of the differences between the average vortex number density for different values of $\tau$, $n_V(t,\tau_1=0.1) - n_V(t,\tau_2)$ vs. time for $\tau_2 = $ 0.5 (Blue), 1.0 (Red), 5.0 (Purple), 10.0 (Green). The black dashed line shows the exhibited power law, \emph{i.e.} $t^{-2}$.}
 \label{vort_difference}
\end{figure}

In Fig.~\ref{vort_nKvstloglog} we show the average vortex number density for different quench parameters 
$\tau$ as a function of time. 
As in the kink case, the plots of  ${n}_V$ vs. $t$ for different $\tau$ converge 
to the same function and decay as $t^{-1}$ as we expect from the 
analytical estimate in \eqref{vortanalres}. The result also agrees with the intuition that a vortex corresponds to the intersection of two independent domain walls.

Fig.~\ref{vort_nKvstloglog} also shows that immediately after the phase transition, $n_V$ increases 
from zero to some maximum value $(n_V)_{\rmmax}$ in a time $t_{\rmmax}$. As time goes 
on $ n_V$ starts to decay. At very early times,
after the phase transition, randomly distributed vortices of positive and negative winding number
are produced, but then the system starts relaxing, the vortices-antivortices start annihilating, and
the dynamics reaches its  scaling regime.

We can also plot the differences of vortex 
number densities for different values  of $\tau$ as we did in the case of kinks: 
$\Delta n_V(t,\tau_1,\tau_2) \equiv n_V(t,\tau_1) - n_V(t,\tau_2)$. 
This is shown in Fig.~\ref{vort_difference} which shows that $\Delta n_V(t,\tau_1,\tau_2)$ decays as
$t^{-2}$ at late times. We thus deduce the asymptotic form,
\be
\label{vort_mainres}
n_V (t) = C_V \Big( \frac{m}{t} \Big) + \mathcal{O}\bigg(t^{-2}\bigg)\,,
\ee
where, $C_V$ is some constant of proportionality which is independent of the quench time 
scale $\tau$. Numerically, we find $C_{V} \approx 0.092$. This is again in reasonable agreement 
with the value we calculated analytically for a sudden phase transition ($\tau = 0$) in 
Eq.~\eqref{vortanalres}, more precisely $
1/\pi^{2}\approx 0.101$.

\begin{figure}[ht]
      \includegraphics[width=0.45\textwidth,angle=0]{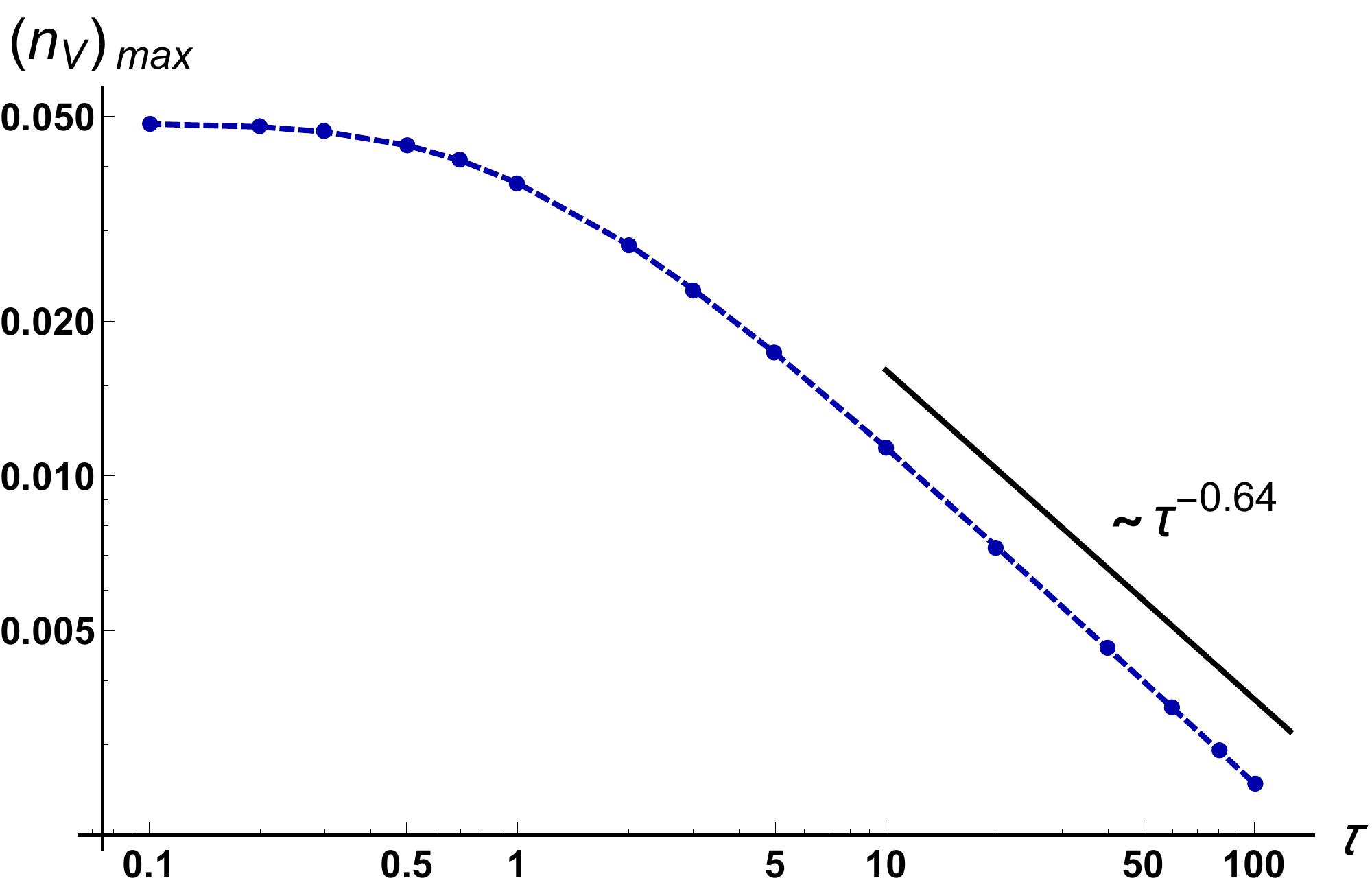}
  \caption{Log-Log plot of the maximum average vortex number density $(n_V)_{\rm max}$ vs. $\tau$. For larger values of $\tau$ the power law manifested is $\sim \tau^{-0.64}$.}
  \label{vort_maxnkvstau}
\end{figure}

\begin{figure}
      \includegraphics[width=0.45\textwidth,angle=0]{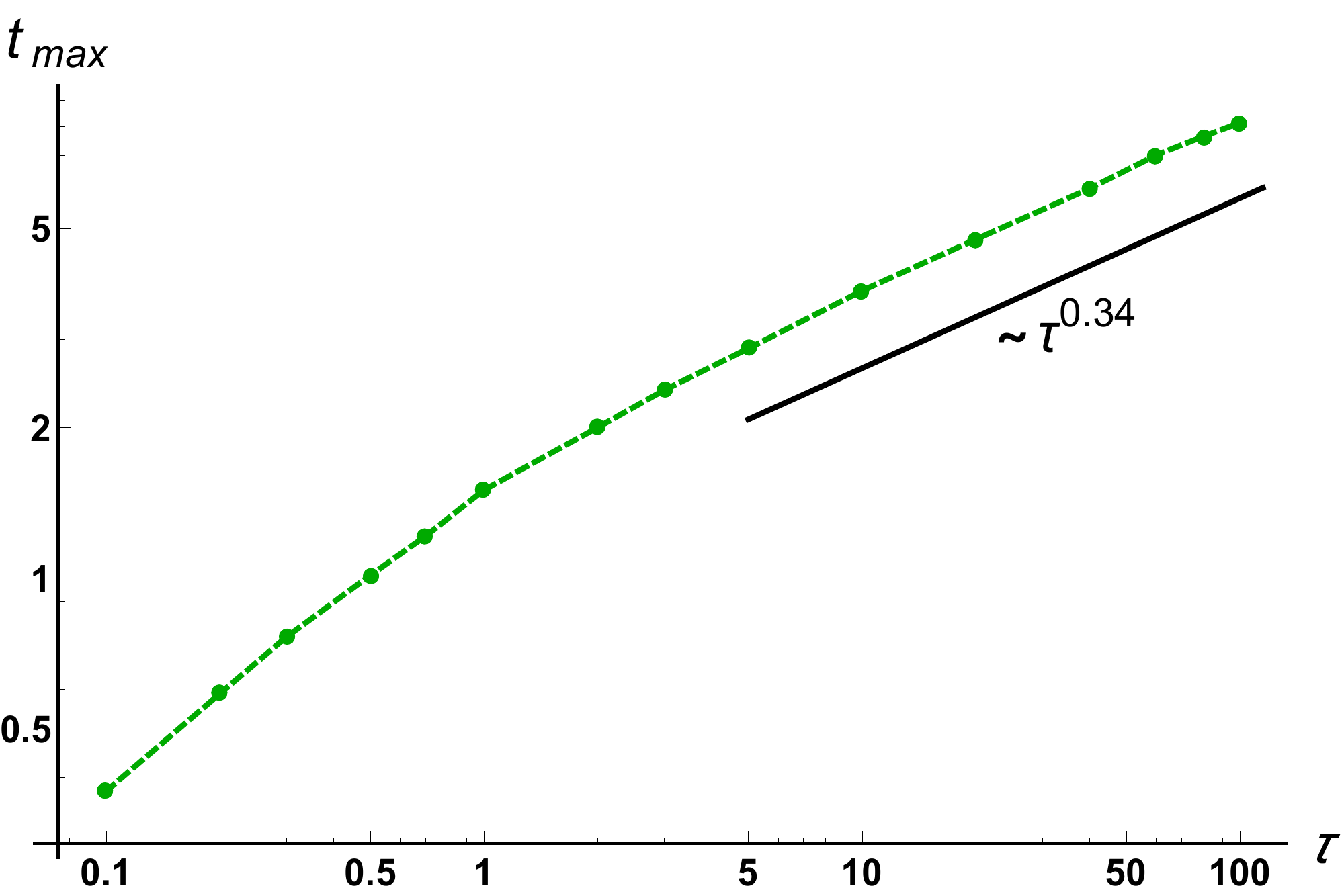}
  \caption{Log-Log plot of the time at which maximum average vortex number density $(n_V)_{\rm max}$ occurs ($t_{\rm max}$) vs. $\tau$. For larger values of $\tau$ the power law exhibited is $\sim \tau^{0.34}$.}
  \label{vort_maxnkt}
\end{figure}

The plots of $(n_V)_{\rmmax}$
versus $\tau$, and $t_{\rmmax}$ versus $\tau$ are shown in Fig.~\ref{vort_maxnkvstau} and Fig.~\ref{vort_maxnkt} respectively. The 
intuitive understanding that a faster phase transition (smaller quench time scale $\tau$) leads to 
greater and more rapid vortex production
is confirmed by these plots. 
Moreover, from Fig.~\ref{vort_maxnkvstau} we see that the maximum number density of vortices 
$(n_V)_{\rmmax}$ flattens as the quench time scale $\tau$ approaches zero. This happens for 
a value $(n_V)_{\rmmax}\approx 0.0483$ which is once again in good agreement with our analytical 
result in Eq.~\eqref{maxnvanal}. 

As a final remark, comparing Fig.~\ref{maxnkvstau} to Fig.~\ref{vort_maxnkvstau} shows us right away that 
for the same quench time-scales $\tau$, the maximum vortex number density $(n_V)_{\rmmax}$ is 
much lower than the maximum kink number density $(n_K)_{\rmmax}$. For example, in the limiting 
case of $\tau \rightarrow 0$, $(n_V)_{\rmmax}\approx 0.050$ while $(n_K)_{\rmmax}\approx 0.175$. 
This is again to be expected since the formation of a vortex requires the simultaneous vanishing 
of two fields, which is less probable than the vanishing of a single field necessary for the formation 
of a kink in one dimension. 

\section{Higher Dimensions: Monopoles}
\label{monopoles}

Having worked out the details of the $d=1$ and $d=2$ cases, it is easy to see that the methods 
described in the previous sections directly generalize to higher dimensions. Without going into 
the details of a rigorous proof, 
the average number density of zero-dimensional topological defects $n_D$ formed in the $d$ dimensional field 
theory discussed in Sec.~\ref{introduction} is given by
\be
 n_D = d! n_K^d = \frac{d!}{2^{d/2}\pi^d} \left ( \frac{m}{t} \right )^{d/2} 
+ {\cal O} \left ( t^{-(d+2)/2} \right )
\label{nD}
\ee
for late times.
The factor of $d!$ arises because of permutation symmetry. To get a monopole in $d$ dimensions
we need coincident zeros of $d$ fields in a cell of the lattice. As in Sec.~\ref{sec:vortices}, the
point $\Phi=0$ corresponds to the intersection of $d$ orthogonal domain walls. The $d!$ permutations
of the wall positions preserves the $\Phi=0$ point which leads to the $d!$ prefactor in
\eqref{nD}.

\begin{figure}[ht]
      \includegraphics[width=0.45\textwidth,angle=0]{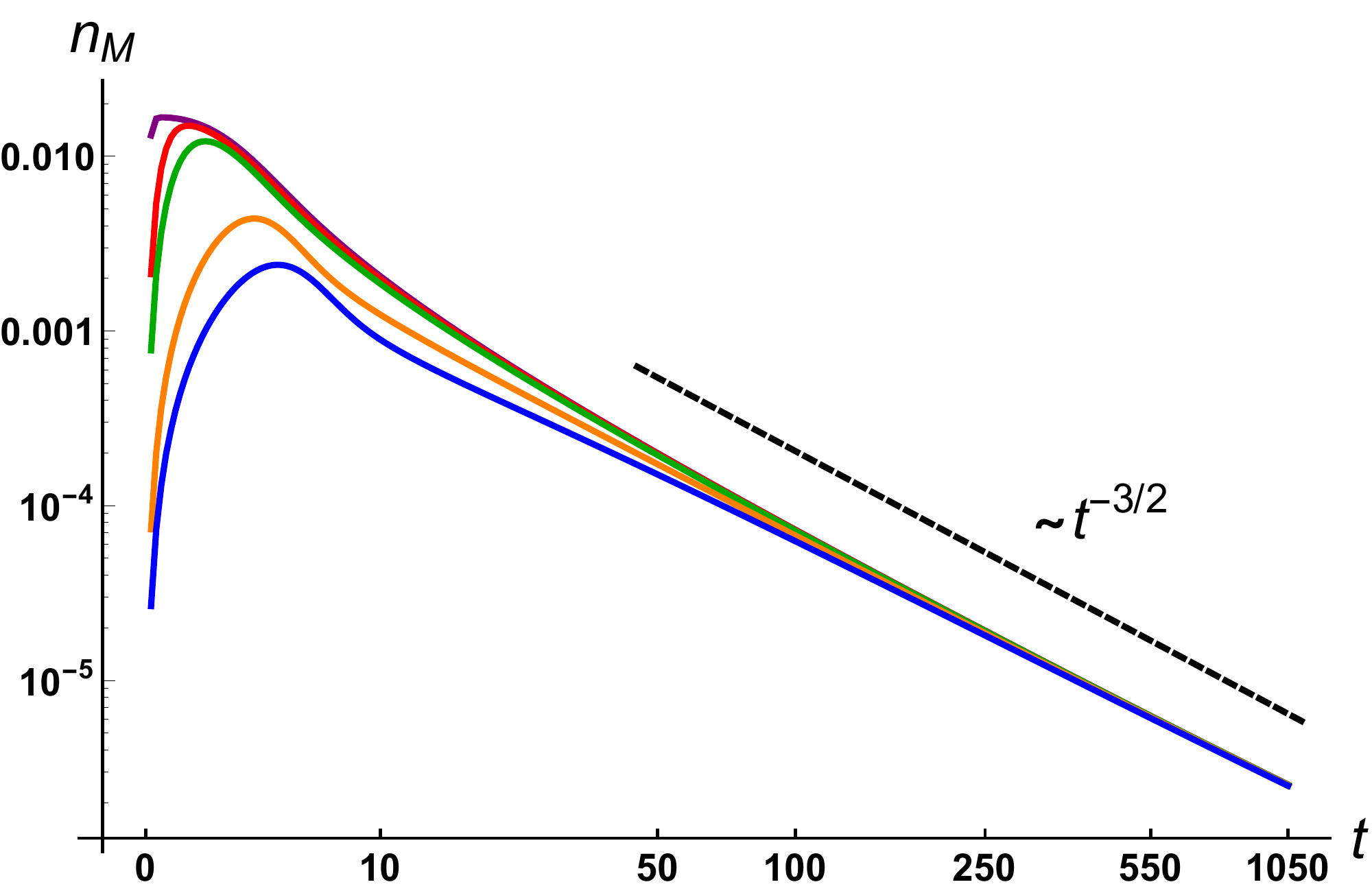}
  \caption{Log-log plot of $n_M (t)$ versus time for $\tau$ =0.1 (Purple, topmost curve), 0.5 (Red), 
  1.0 (Green), 5.0 (Orange), 10.0 (Blue). The black dashed line shows the exhibited power law at 
  late times, \emph{i.e.} $t^{-3/2}$. Here we use $L = 800$, $N = 1600$.}
  \label{mono_nKvstloglog}
\end{figure}

\begin{figure}[ht]
      \includegraphics[width=0.45\textwidth,angle=0]{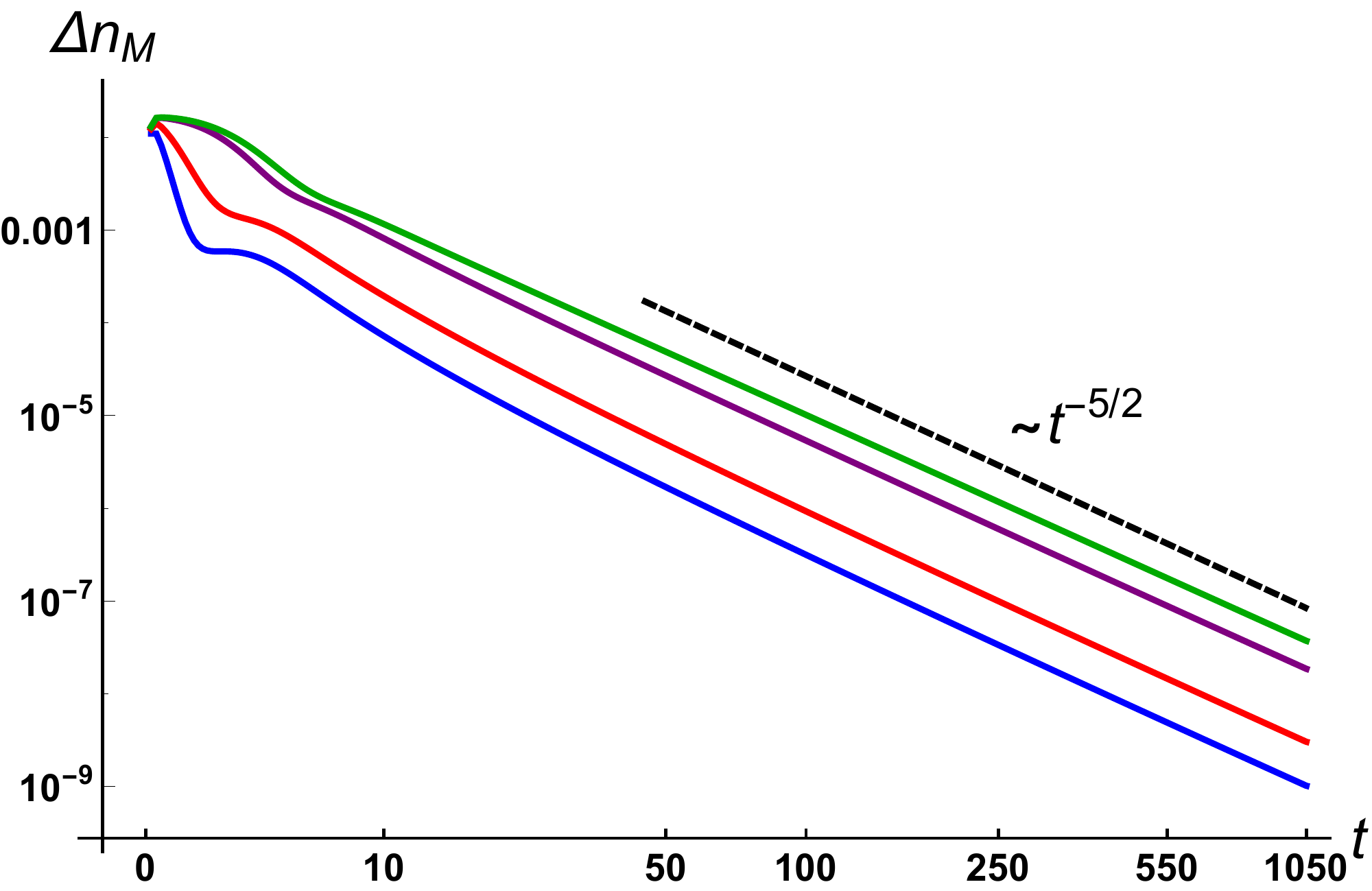}
  \caption{Log-log plot of the differences between the average monopole number density for different values of $\tau$, 
  $n_M(t,\tau_1 = 0.1) - n_M(t,\tau_2)$ vs. time for $\tau_2 = $ 0.5 (Blue), 1.0 (Red), 
  5.0 (Purple), 10.0 (Green). The black dashed line shows the exhibited power law, \emph{i.e.} $t^{-5/2}$.}
  \label{mono_difference}
\end{figure}

In Fig.~\ref{mono_nKvstloglog} we show numerical results for $d=3$ for the monopole number
density as a function of time, obtaining the first term on the right-hand side of \eqref{nD}. 
In Fig.~\ref{mono_difference} we provide evidence for the second term on the right-hand
side of \eqref{nD}.

\section{Effect of self-interactions}
\label{discussion}


A key question is to understand the range of parameters for which our results
are a good approximation even when $\lambda\ne 0$. We will address this in the context of 
the model in one spatial dimension given in \eqref{kinkqft}. We check for self-consistency
of our solution and examine the conditions under which it breaks down.
%
%

\begin{figure}[ht]
      \includegraphics[width=0.45\textwidth,angle=0]{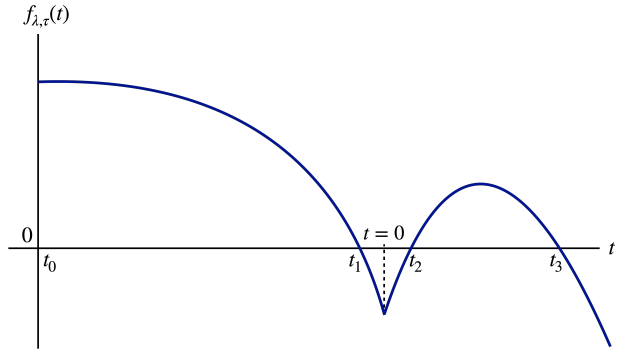}
  \caption{
Sketch of $f_{\lambda,\tau}(t)$ to show its generic features.
  }
  \label{diffplot}
\end{figure}

Our solution for the wavefunction is a Gaussian at all times and so 
$\la \phi^4 \ra = 3 \la \phi^2 \ra^2$.
With $\lambda \ne 0$, the evolution of the wavefunctional, $\Psi[\phi,t]$, will depend on $\lambda$. 
As long as $\Psi$ can be approximated by a Gaussian centered at $\phi=0$ 
we can use the Hartree approximation ({\it e.g.}~\cite{baym2018lectures})
to write $\lambda \phi^4$ as $3 \lambda \la \phi^2 \ra \phi^2$. 
Taking into account mass renormalization at lowest order in $\lambda$ we obtain an effective mass squared 
$m^{ \rm eff}_{2}(t)$,
\be
\label{effmass}
m^{\rm eff}_{2}(t) = m_{2}(t)+\frac{3}{2} \lambda \la \phi^2 \ra_{\rm in} -\frac{3}{2} \lambda \la \phi^2 \ra\,.
\ee
where the ``in'' subscript refers to evaluation at the initial time ($t \to -\infty$).
The mass counterterm $3\lambda \la \phi^2 \ra_{\rm in}/2$ is chosen such that the effective mass equals 
$m$ at the initial time.
Therefore, in the Hartree approximation, the effects of interactions are negligible if the $\lambda$ dependent
corrections to $m_2$ are small and $m^{ \rm eff}_{2}(t)\approx m_2(t)$, or, 
\be
3 \lambda \left[ \la \phi^2 \ra - \la \phi^2 \ra_{\rm in} \right] \ll 2|m_2|\,.
\label{condition}
\ee
The condition in \eqref{condition} will fail in two circumstances. First, 
around the time of the phase transition, $t\sim 0$, when $m_2 \sim -m^2 t/\tau$ (see \eqref{m2t}) is very 
small; second, at late times, when $\la\phi^2\ra$ grows large. We can make these statements more precise 
by noticing that \eqref{condition} is strongly violated whenever the function 
\be
f_{\lambda,\tau}(t)\equiv  2|m_2|- 3 \lambda \left[ \la \phi^2 \ra - \la \phi^2 \ra_{\rm in} \right]
\ee
becomes negative. It turns out that, generically, $f_{\lambda,\tau}$ has three zeros that we denote as
$t_1$, $t_2$, $t_3$ and 
it is negative on the intervals $[t_1,t_2]$ and $[t_3,\infty)$ (see Fig.~\ref{diffplot} for a qualitative sketch of 
$f_{\lambda,\tau}$). The late time violation is not important for us as long as by that time all the kinks have 
already been formed. Moreover their mutual interactions  are exponentially suppressed on distances longer 
than $1/m$ in $d=1$, and they can be completely neglected given that the average separation of the kinks 
is larger than $\sim (n_{K})_{\rm max}^{-1}\sim 6/m$. On the other hand, the early time violation
in the interval $[t_1,t_2]$ can be 
important as it might interfere with kink production and change the maximum kink number density. 

We can thus deduce three necessary conditions for the kink number density in the $\lambda=0$ model 
to be a good approximation to that in the $\lambda\neq 0$ case:
\begin{enumerate}
\item[(i)] The duration of early time violation of \eqref{condition} needs to be finite {\it i.e.} $t_2<\infty$.
\item[(ii)] All the kinks need to have been produced by the time the late time violation of \eqref{condition} 
sets in {\it i.e.} $t_{\rm max}<t_3$
\item[(iii)] The duration of the early time violation of \eqref{condition} needs to be much smaller than the 
fastest timescales of variation of the wave-functional {\it i.e.} $\Delta t\equiv t_2-t_1\ll 1/m$.
\end{enumerate}

\begin{figure}[ht]
      \includegraphics[width=0.47\textwidth,angle=0]{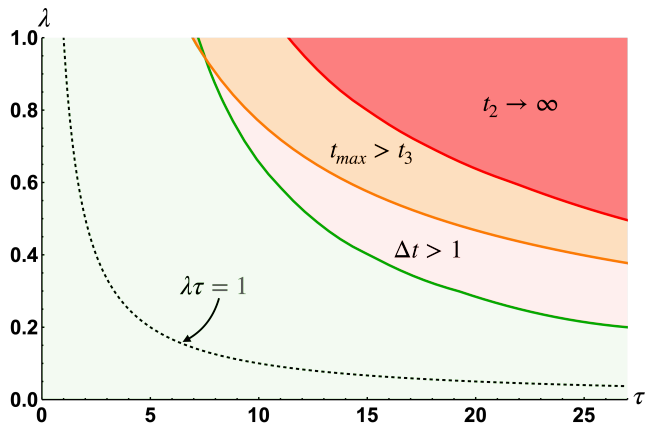}
  \caption{Plot showing the allowed and disallowed regions of the $(\lambda,\tau)$ parameter space in units where $m=1$.}
  \label{region}
\end{figure}

We have swept the $(\lambda, \tau)$ parameter space to determine the regions where the above conditions are verified. This has been done numerically by approximating $f_{\lambda,\tau}$ via
\be
f_{\lambda,\tau}(t) \approx  2|m_2|- 3 \lambda \sum_{n=1}^N\left(|c_n(t)|^2 - |c_n(t_0)|^2 \right)\,,
\ee
and determining the corresponding values of $t_1$, $t_2$, $t_3$ for a wide range of values of $\lambda$ and $\tau$. The results are shown in Fig.~\ref{region} where we used the same numerical parameters as in Sec.~\ref{numerics}. The regions shaded in red, orange and pink are excluded by the necessary conditions (i), (ii) and (iii) respectively. Alternatively we expect the $\lambda=0$ model to be accurate inside the green region. Remarkably, the $\lambda\tau/m =1$ curve lies deep inside this region which indicates that $\lambda\tau/m\ll 1$ is a sufficient condition for the approximation to be valid.



\section{Conclusions}
\label{conclusions}

In this work we carried out a thorough analysis of the dynamics of topological defect formation in a quantum field theory where the only interactions are with external parameters
that induce a quantum phase transition. We thus worked in the limit where self-interactions can be neglected. Results for the number density of kinks
in one spatial dimension are summarized in Fig.~\ref{nKvstloglog}, for vortices in two
spatial dimensions in Fig.~\ref{vort_nKvstloglog}, and for monopoles in three spatial dimensions
in Fig.~\ref{mono_nKvstloglog}. These results indicate that the number density of topological defects in $d$ spatial dimensions scales as $t^{-d/2}$ and does not depend on the quench time scale, in the late time limit. Moreover, we showed that the sudden phase transition analytical result is a universal attractor. These novel results stand in contrast to the Kibble-Zurek
prediction for a thermal phase transition.

We have also discussed the limit within which our results can be expected
to be a good approximation for a more realistic theory where self-interactions are not explicitly set to zero. In the case of kinks ($d=1$) we found the condition $\lambda \tau/m \ll 1$ where $\lambda$ is the self-interaction coupling 
strength, to be a sufficient condition for our results to hold. This condition 
can be generalized on dimensional grounds to be $\lambda m^{d-2} \tau \ll 1$ in $d$ 
spatial dimensions.

\acknowledgements
We thank Dan Boyanovsky for comments. Computations for this work were performed on the Agave cluster at Arizona State University. MM is supported by the National Science Foundation grant numbers PHY-1613708 and PHY-2012195. 
TV is supported by the U.S.  Department of Energy, Office of High Energy Physics, 
under Award No.~DE-SC0019470 at Arizona State University. GZ is supported by 
{\it Moogsoft} and the Foundational Questions Institute (FQXi). 

\appendix

\bibstyle{aps}
\bibliography{long_paper}

\end{document}